\documentclass[a4paper,twocolumn,aps,superscriptaddress,floatfix,prmaterials]{revtex4-2}
\usepackage{graphicx}
\usepackage{comment}
\usepackage{amsmath,amsfonts,amssymb,amsthm}
\usepackage{mathtools,braket}
\usepackage{url}
\usepackage{siunitx}
\usepackage{hyperref}
\usepackage{xcolor}

\newcommand{\ba}{\begin{align}}
\newcommand{\ea}{\end{align}}
\newcommand{\be}{\begin{equation}}
\newcommand{\ee}{\end{equation}}
\newcommand{\bml}{\begin{multline}}
\newcommand{\eml}{\end{multline}}
\newcommand{\bea}{\begin{eqnarray}}
\newcommand{\eea}{\end{eqnarray}}

\def\d{\delta}

\def\l{\lambda}

\def\w{\omega}

\def\blk{{\mathbf k}}

\def\blq{{\mathbf q}}

\def\blP{{\mathbf P}}

\newcommand{\ai}{ab--initio}

\begin{document}

\title{Excitons and carriers in transient absorption and time--resolved ARPES spectroscopy: an abinitio approach}
\author{D. Sangalli}
\affiliation{Istituto di Struttura della Materia of the National Research
Council, Via Salaria Km 29.3, I-00016 Montelibretti, Italy}
\date{\today}

\begin{abstract}
I present a fully \ai\, scheme to model transient spectroscopy signals in presence of strongly bound excitons. Using LiF as a prototype material, I show that the scheme is able to capture the exciton signature both in time-resolved ARPES and transient absorption experiments. The approach is completely general and can become the reference scheme for modelling pump and probe experiment in a wide range of materials.
\end{abstract}

\maketitle

\section*{Introduction}

The exciton is a quasi--particle composed by an electron--hole pair bound via Coulomb interaction~\cite{Mahan1967}.
In condensed matter physics, exciton energies, $\omega_{\lambda\blq}$, and wave--functions, $A^{\lambda\blq}$, are obtained via the solution of the \ai\, Bethe--Salpeter equation (BSE), and well describe the neutral excitations of materials in the linear response regime~\cite{Onida2002,Albrecht1998}. Indeed the energy difference between the electronic ground state, $|\Psi_0^N\rangle$, and the neutral excited state, $|\Psi_I^N\rangle$, is well approximated by $\omega_{\lambda\blq}\approx(E_I^N-E_0^N)$. Similarly the many--body wave--function can be approximated as $|\Psi_I^N\rangle \approx \hat{e}^\dag_{\lambda\blq} |\Psi_0^N\rangle$, via the definition of the exciton creation operator $\hat{e}^\dag_{\lambda\blq}=\sum_{cv\blk} A^{\lambda\blq}_{cv\blk} \hat{a}^\dag_{c\blk}\hat{a}^\dag_{v\blk-\blq}$.
Neutral excited states can be generated via the use of optical perturbations, such as laser pulses, if the frequency of the pulse is tuned in resonance to the absorption peaks in the bound region of the spectrum of a material.

The interest in the physics of the excitons grew significantly in recent years, for both fundamental and technological reasons. On one side, the development of femto--second laser--pulses made it possible to explore ultra-fast electron dynamics, with the exciton playing a key role due to the optical nature of the perturbation~\cite{Sundaram2002,Manzoni2015,Koch2006,Chemla2001}. Different works have discussed the possible generation of excitonic states and in particular of nonequilibrium Bose--Einstein condensate (BEC) of excitons ~\cite{Schmitt-Rink1988,Moskalenko2000,Perfetto2019b,Rustagi2019,Christiansen2019,Triola2017}. On the technological side, the growing interest in materials for applications to photovoltaic devices and semiconductor devices, calls for accurate modelling of the absorption process and the interaction between electrons and holes which, in turn, strongly affects carriers mobility. Many new materials, in particular layered materials such as transition metal dicalcogenides~\cite{Qiu2013a,MolinaSanchez2013,Manca2017}, and chromium or bismuth trihalide~\cite{MolinaSanchez2020,Mor2021}, display remarkably high binding energy for semiconductors. When studied through pump and probe experiments, such materials are often driven via laser pulses tuned resonant with such excitonic energies. The transient absorption (TR-abs) or the time--resolved angle--resolved photo--emission (TR-AR-PES) signals are then detected~\cite{Stolow2004,Ueba2007,Bovensiepen2012,Smallwood2016,Freericks2009}. It is then crucial to have accurate modelling and understanding of both the state generated by the pump and the signal measured by the probe.

Due to the coherent nature of the laser pulse, the system is sent into coherent superpositions of the ground state and neutral excitations, which is usually referred to as ``coherent excitons''. 
Many works in the literature have been focused on the description of such experiments~\cite{Henneberger1986a,Kappei2005,Berghauser2018} modelling coherent excitons, either as an ideal boson or via a parametric Wannier equation~\cite{Trovatello2020,Christiansen2019,Malic2018}. On the other hand in the \ai\, community the modelling of such experiments have focused on a detailed description of the material properties and the photo--excitation process, but somehow neglecting the role of the electron--hole interaction and thus describing the generated nonequilibrium density in terms of free carriers.
The main reason behind this is the intrinsic difficulty in a fully \ai\, modelling of excitons in the nonequilibrium regime. Time dependent density functional theory (TDDFT), the workhorse for \ai\, nonequilibrium dynamics~\cite{DeGiovannini2013,DeGiovannini2017}, cannot easily capture the physics of the exciton~\cite{Reining2002,Marini2003}. The recent attempts to go beyond TDDFT, within \ai\, non--equilibrium (NEQ) Many Body Perturbation Theory (MBPT) have, so far, not accounted for the physics of the exciton in the description of the probed signal~\cite{Pogna2016,Smejkal2021,Wang2018,MolinaSanchez2017}.
As a consequence the interpretation of experimental signal has seen the use of different concepts, sometimes reaching completely different conclusions~\cite{Pogna2016,Smejkal2021,Trovatello2020,Berghauser2018,Wang2018}.

The goal of the present work is to close this gap and model fully \ai\, the transient signal, taking into account the role of the electron--hole interaction in both the pumping process and the probe process. I will use LiF as a prototype material which displays a strongly bound exciton~\cite{Marini2004}, and consider the physical regime where (i) the pump and the probe do not overlap in time, and (ii) the pump pulse is tuned resonant to the absorption of LiF, thus generating a real NEQ population. Moreover, a low pump pulse intensity is used to remain below the exciton Mott transition~\cite{Koch2003,Perfetto2020}. We emphasize that the scheme used in the present work could be also employed in other physical regimes, for example, to describe the overlapping regime, where the dynamical Stark effect plays a key role~\cite{Schmitt-Rink1986,Schmitt-Rink1988}, and in the case where the pump is tuned off-resonance (i.e. with a frequency below the optical gap) leading to the generation of a transient virtual NEQ population and of the Floquet replica~\cite{Giovannini2019}. This is however beyond the goal of the present work.

\section{Coherent and non-coherent states}

\begin{figure}[t]
\begin{center}
\includegraphics[width=0.45\textwidth]{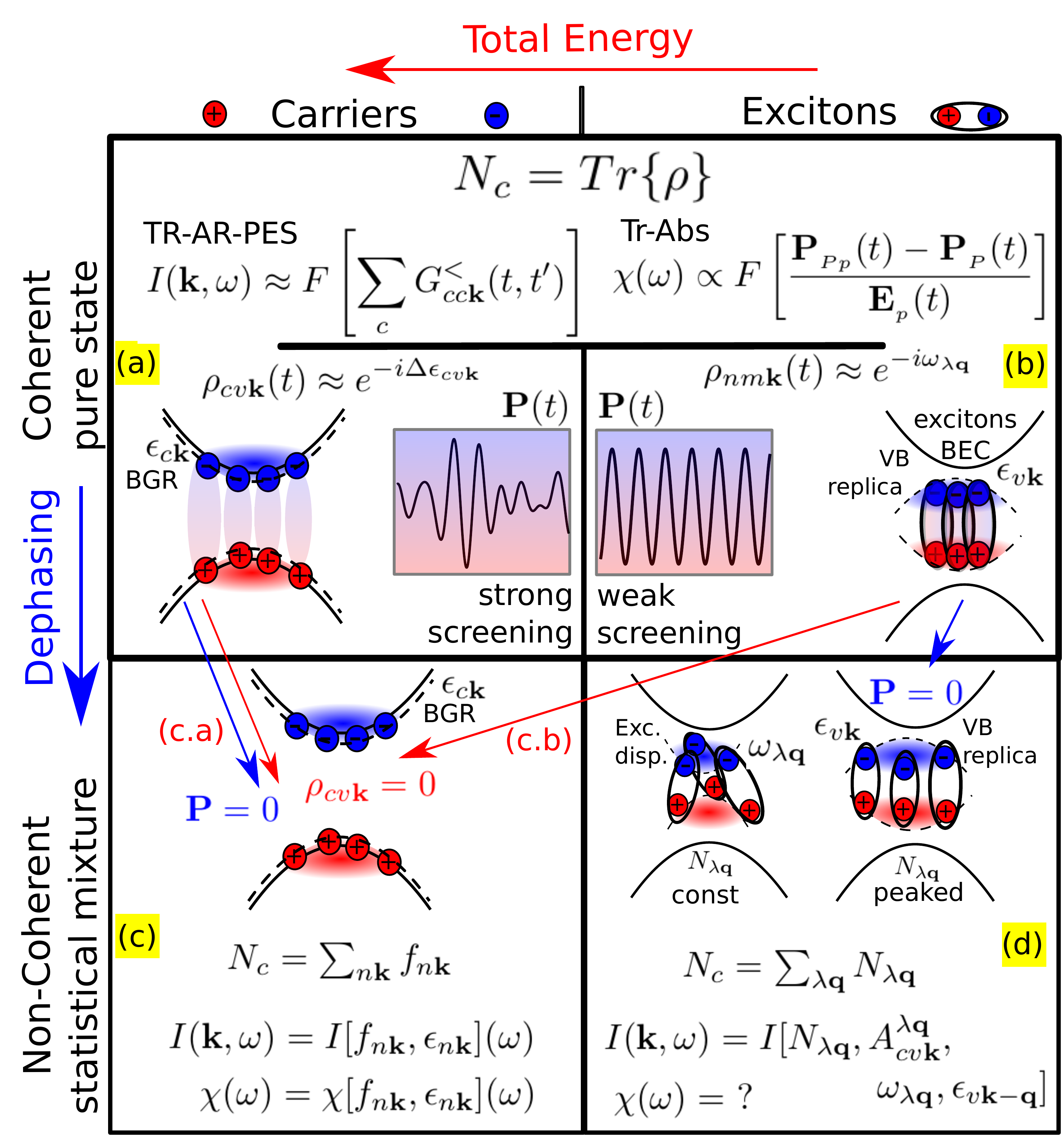}
\end{center}
\caption{Overview of the four non--equilibrium states considered in the present work. They are divided in free carriers and bound exciton states (left to right), and in coherent pure states and non-coherent statistical mixtures (top to bottom). $N_c$ is the excited density in the conduction band. The equation used to reconstruct the TR-ARPES spectral function $I(\blk,\omega)$ and the response function $\chi(\omega)$ are scketched in the top. In the bottom the approximation used in the literature and holding for non--coherent case are also reported. The graphical representation of the states is realized via their TR-ARPES signal.}
\label{fig:neq_states_overview}
\end{figure}
In Fig.~\ref{fig:neq_states_overview} an overview of the physical states considered in the present work is shown. I'll use the labels of the panels in this figure to identify these states in the manuscript: (a) for coherent carriers, (b) for coherent excitons, (c.a) for non--coherent carriers obtained dephasing the coherent carriers state, (c.b) for non--coherent carriers obtained via artificial dephasing of the coherent exciton state, and (d) for non--coherent excitons obtained via physical dephasing of the excitonic state. Physical and artificial dephasing are represented via blue and red lines respectively in the figure. Their meaning will be discussed in the manuscript. The two processes are identical and lead to the same state in presence of free carriers, while they must be distinguished and lead to different states in presence of bound excitons.

The states are generated via the propagation of the equation of motion (EOM) for the time--dependent density matrix, $\rho_{nm\blk}(t)$ in presence of a pump field. The hartree plus screened exchange approximation (HSEX), with the screening kept frozen at equilibrium, is used for the many-body self-energy entering the EOM~\cite{Attaccalite2011}. No further approximation is introduced.
\begin{equation}
i\partial_t\rho_{nm\blk}=
\Delta\epsilon_{nm\blk}\rho_{nm\blk}+
[\Delta\Sigma^{HSEX},\rho]_{nm\blk}+
[U^{ext},\rho]_{nm\blk}
\label{eq:EOM_rho}
\end{equation}
Starting from $\rho_{nm\blk}(t)$, I reconstruct the TR-ARPES and the Tr-Abs signal.
The TR-ARPES spectral function $I(\blk,\omega)$ in the conduction region is obtained via a Fourier transform of $G_{cc'\blk}^<(t,t')$~\cite{Perfetto2016}, with $G_{nm\blk}^<(t,t')$ reconstruct from $\rho_{nm\blk}(t)$ using the generalized Kadanoff Baym ansatz, which is exact within the chosen approximation. The Tr-Abs signal $\chi(\omega)$ is extracted via the Fourier transform of the time dependent polarization, $\blP(t)=\blP_{Pp}(t)-\blP_{P}(t)$ obtained performing one simulation with only the pump pulse and another with both the pump and the probe pulse, as already discussed in the literature~\cite{Sato2014,Perfetto2015}. The signal depends on the probe pulse delay $\tau$ from the pump pulse. However, the $\tau$ dependence in the equations is omitted here since I focus on the transient signal due to the states generated right after the action of the pump pulse. The signal extracted from such coherent states will be compared with the signal generated from a non--coherent population of carriers $f_{n\blk}$ or excitons $N_{\lambda\blq}$. The update of the screening is neglected also in the definition of the transient signal. The reason of this choice is twofold. (i) We want to stick into the HSEX approximation with equilibrium screening in all the steps. (ii) We aim to a clear comparison between the signal generated in the excitonic case, with the signal generated in the carriers' case. In the excitonic case, the update of the screening is expected to be negligible in the low--density regime~\cite{Perfetto2020b,Moskalenko2000} i.e. below the exciton Mott transition density. Above such density excitons evaporate, going through a BCS like regime of electrons and holes~\cite{Perfetto2019b,Kappei2005} and it is not meaningful anymore to speak about ``excitonic case''. In the case of free carriers there exist also a Mott transition density (this has been studied in doped silicon for example~\cite{Jain1976}) above which the carriers become a metallic gas and the screening update could easily become the dominant contribution, washing out a possible detailed comparison of other effects. Thus the results presented here are meaningful for densities below both such Mott transitions.

\begin{figure}[t]
	\begin{center}
		\includegraphics[width=0.450\textwidth]{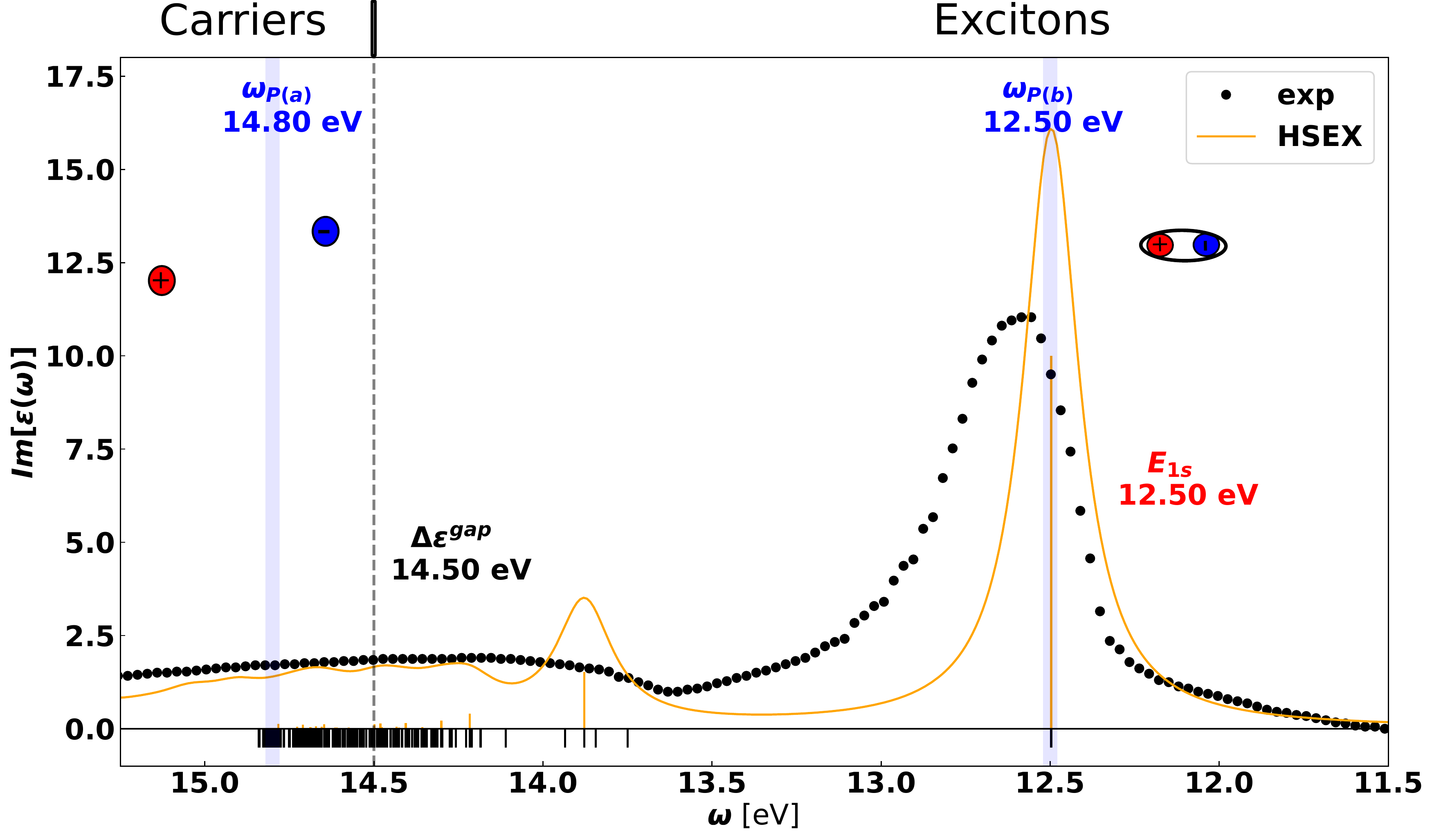}
	\end{center}
	\caption{The optical absorption of LiF. Vertical lines mark the position of the bright eigenvalues of the BSE equation weighted by the assocaited oscillator strength in the positive $y$-axis. The full BSE spectrum is represented in the negative  $y$-axis. The vertical dashed lines represents the position of the electron-hole continuum, i.e. the GW band gap. All the BSE eigenvalues at lower energy are part of the Rydberg exciton series.}
	\label{fig:absorption}
\end{figure}
As shown in Fig.~\ref{fig:absorption}, LiF absorption is dominated by a peak of excitonic nature at about 12.5~eV with a binding energy of almost 2~eV.
Two possible frequencies, ${\omega_P}$, are then considered for the pump laser pulse used: (i) ${\omega_{P_a}=14.8}$~eV above the band gap of LiF, and (ii) ${\omega_{P_b}=12.5}$~eV resonant to the lowest excitonic peak of LiF (see fig.~\ref{fig:absorption}). The laser pulse is introduced in the form of an oscillating function centred at the pump energy, multiplied by a Gaussian which determines the pulse duration: $sin(\omega_P t)e^{-(t-t_0)^2/2\sigma^2}$. Here I use $\sigma=10$~fs. 
In the two cases the EOM is propagated for 160~fs, generating the two coherent states of fig.~\ref{fig:neq_states_overview}.
The laser tuned in the continuum generates free carriers with an associated polarization (see fig.~\ref{fig:neq_states_overview}.a). In this regime, the electron--hole interaction has a negligible effect. The propagation at the independent-particles (IP) level is a good approximation to the TD-HSEX scheme, and it overcomes converge issues~\footnote{See Supplemental Material for a detailed comparison of TD-IP and TD-HSEX when the system is driven in the continuum}.
The laser tuned resonant with the excitonic peak generates a coherent excitonic state with an associated polarization (see fig.~\ref{fig:neq_states_overview}.b). 

Dephasing is expected to take place and eventually lead to a statistical mixture of states, both for carriers (fig.~\ref{fig:neq_states_overview}.c) and excitons (fig.~\ref{fig:neq_states_overview}.d). For non interacting electrons, i.e. at the IP level, the most natural way to produce a statistical mixture via the EOM for $\rho(t)$ is to introduce a dephasing term, $-\eta\rho_{nm\blk}$ for $\{nm\}$ such that $\Delta\epsilon_{nm}>\epsilon_{thresh}$. This kills the elements of the density matrix that display coherent oscillations, and it produces a state which is time--independent. The remaining elements of the density matrix can be interpreted as the nonequilibrium electronic occupations: $f_{n\blk}=\rho_{nn\blk}$. The described dephasing procedure corresponds to a real dephasing process in the case of free carriers. Indeed when applied to the coherent carriers state, the IP total energy of the system is conserved.
In presence of interacting electrons instead, such procedure should be performed at the level of the many--body density matrix.
For the excitonic case captured at the HSEX level, interacting electron--hole pairs are considered and dephasing could be introduced at the level of the two--body density matrix~\cite{Sangalli2018a}. Applying the dephasing at the level of the one--body density matrix introduces a change in the total energy of the system, and leads to the formation of carriers populations in place of bound exciton populations. The de-phasing term which needs to be introduced in the EOM for $\rho(t)$ to generate a statistical distribution of excitonic population is however not known.

To model non coherent states, two further time propagation are performed, identical to the previous two, but with also the simple dephasing $-\eta\rho_{nm\blk}$ we just described. Due to the above discussion, with this procedure, two non coherent carriers states are produced. We refer to (c.a) as the state obtained from the dephasing of the coherent carriers state (a), and to (c.b) as the state obtained from the dephasing of the coherent excitons state (b) (see fig.~\ref{fig:neq_states_overview}.c). The dephasing term is switched on after the end of the pump laser pulse, at $t=60$~fs and then switched off at $t=160$~fs, i.e. before the action of the probe pulse. Since the dephasing acts only after the end of the laser pulse, $\rho^{(a/b)}_{nm\blk}= \rho^{(c.a/b)}_{nm\blk}$ for the states with $\Delta\epsilon_{nm}\leq\epsilon_{thresh}$ (here the superscript refer to the state label). 

\begin{figure}[t]
	\begin{center}
		\includegraphics[width=0.450\textwidth]{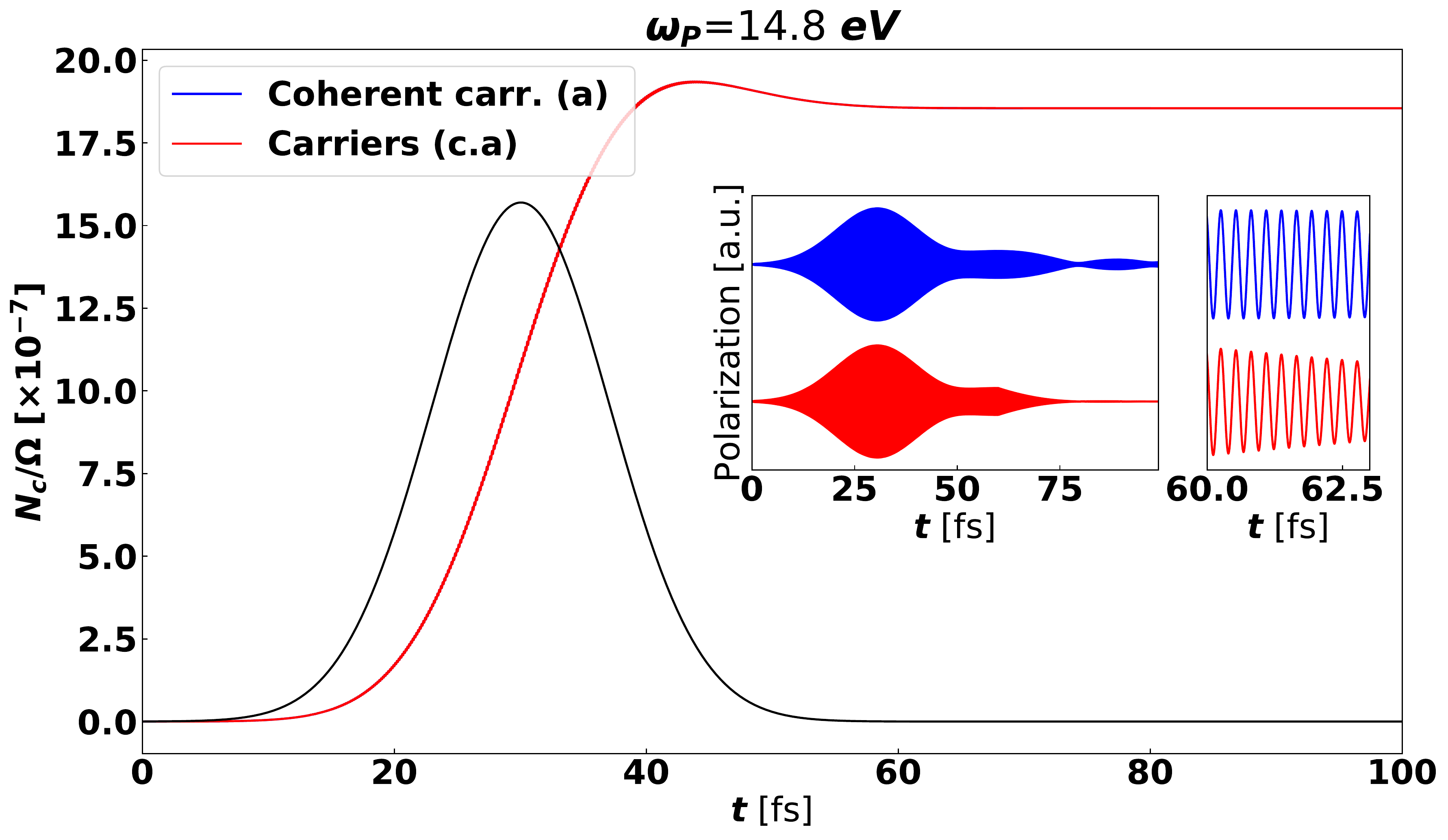}
		\includegraphics[width=0.450\textwidth]{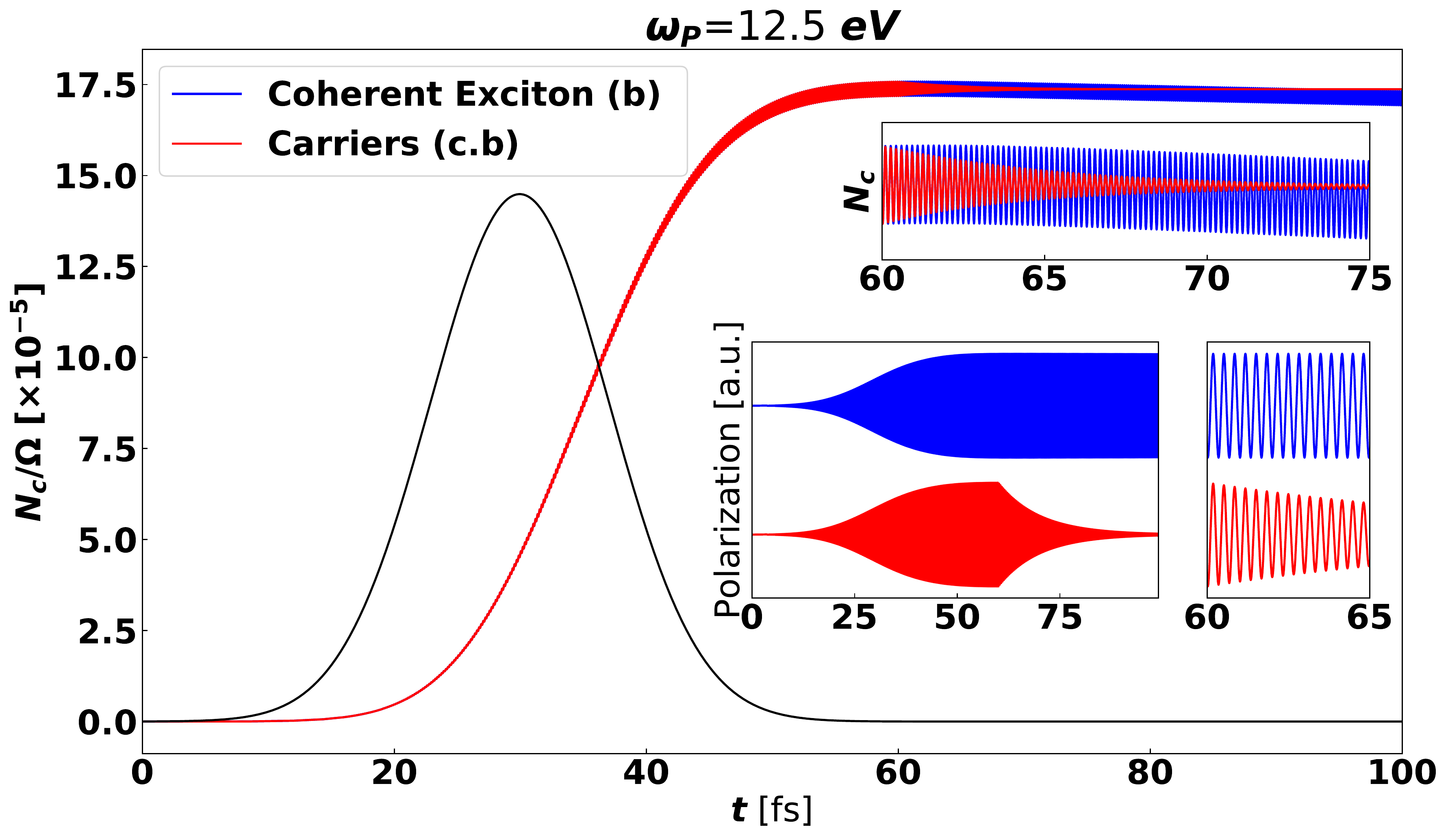}
	\end{center}
	\caption{Time dependent plot of the total carrier density $N_c$ per unit cell (blue and red lines) and laser pulse intensity profile of the pump (black line) are shown in the main frames. Coherent carriers (a) and non--coherent carriers (c.a) are shown in the upper panel. Coherent exciton (b) and non-coherent carriers distribution obtained dephasing the excitonic state (c.b) in the lower panel. This corresponds to four states introduced in Fig.~\ref{fig:neq_states_overview}. A zoom of the carrier density $N_c$ is shown in the uppermost inset for the excitonic case. All other insets show the associated time dependent polarization in different time-ranges.}
	\label{fig:generated_states}
\end{figure}

Thus in summary I have generated four states after 160~fs of simulation and corresponds to the states (a) and (c.a), and (b) and (c.b) for Fig.~\ref{fig:neq_states_overview}. In Fig.~\ref{fig:generated_states} the total excited density $N_c(t)=\sum_{c\blk}\rho_{cc\blk}(t)$ and the polarization $\blP(t)=e\sum_{n\neq m\blk} \rho_{nm\blk}(t)$ of LiF for these four states are shown. The Gaussian envelop of the laser pulse is also shown. 
The polarization of the coherent carriers state contains a continuum of frequencies and experiences free polarization decay (FPD) on a very short time scale. Here the FPD is not complete due to the finite k--points sampling. On the other hand in the coherent excitons case, a single frequency exists, and the polarization is everlasting. This already suggests that the role of the coherent oscillations in the spectrum is expected to be more important in the excitonic case. For each state, I will further propagate the EOM without any dephasing term from $t=160$~fs to $t=260$~fs, first without and later with a probe pulse. A delta function at $t=165$~fs is used for the probe pulse.

\section{Time--Resolved ARPES}

\begin{figure*}[t]
	\begin{center}
	\includegraphics[width=0.32\textwidth]{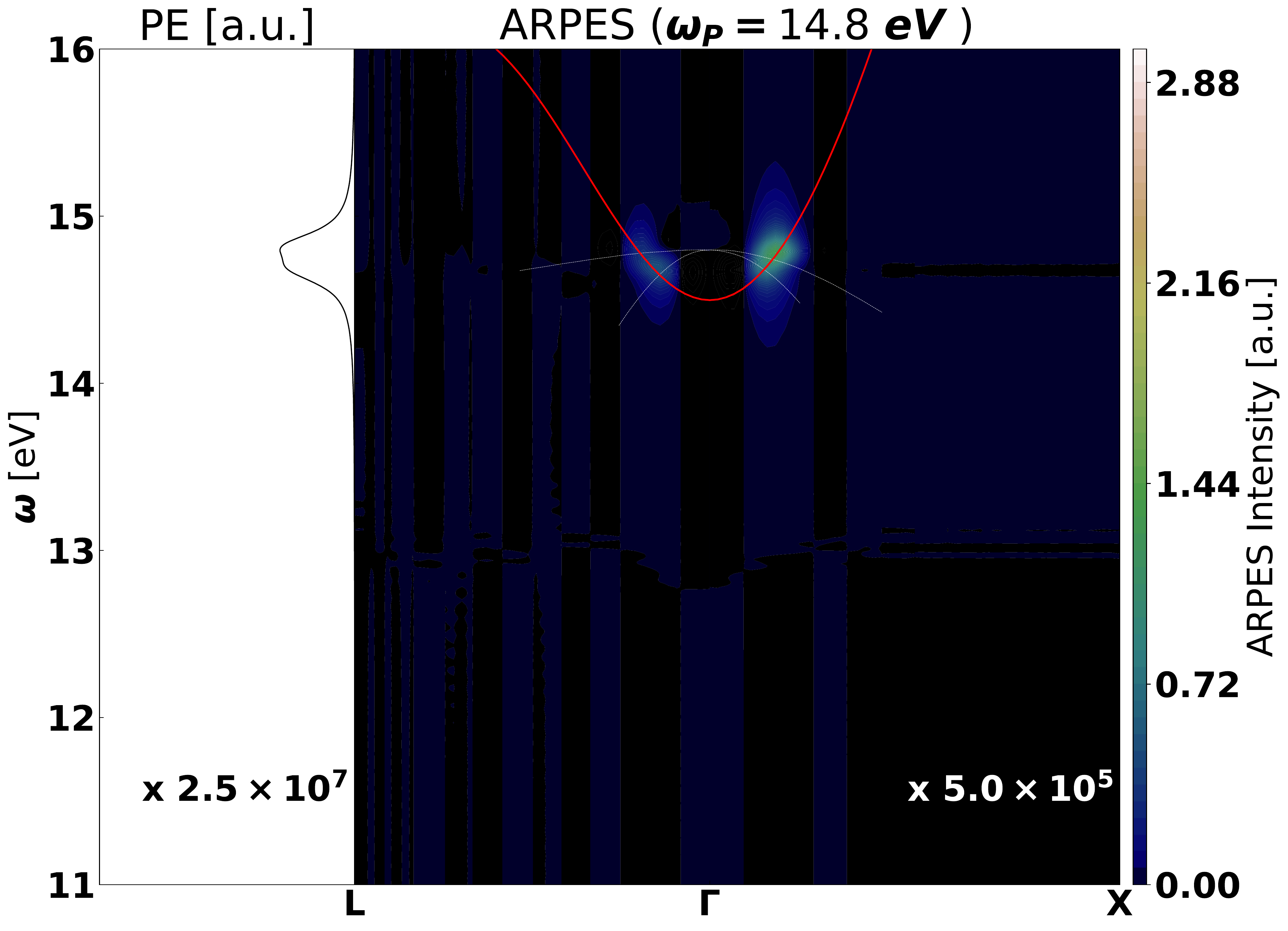}
	\includegraphics[width=0.32\textwidth]{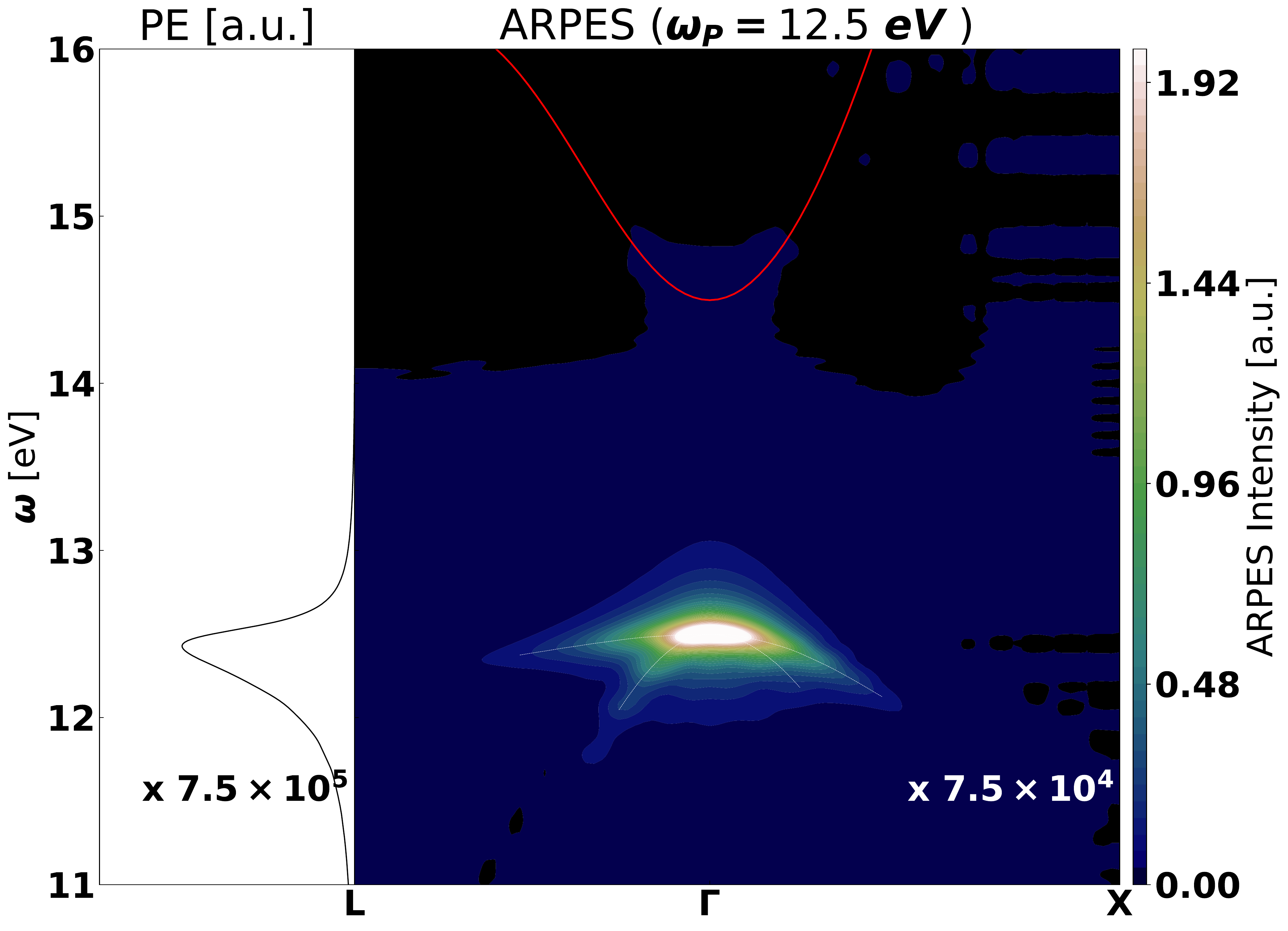}
	\includegraphics[width=0.32\textwidth]{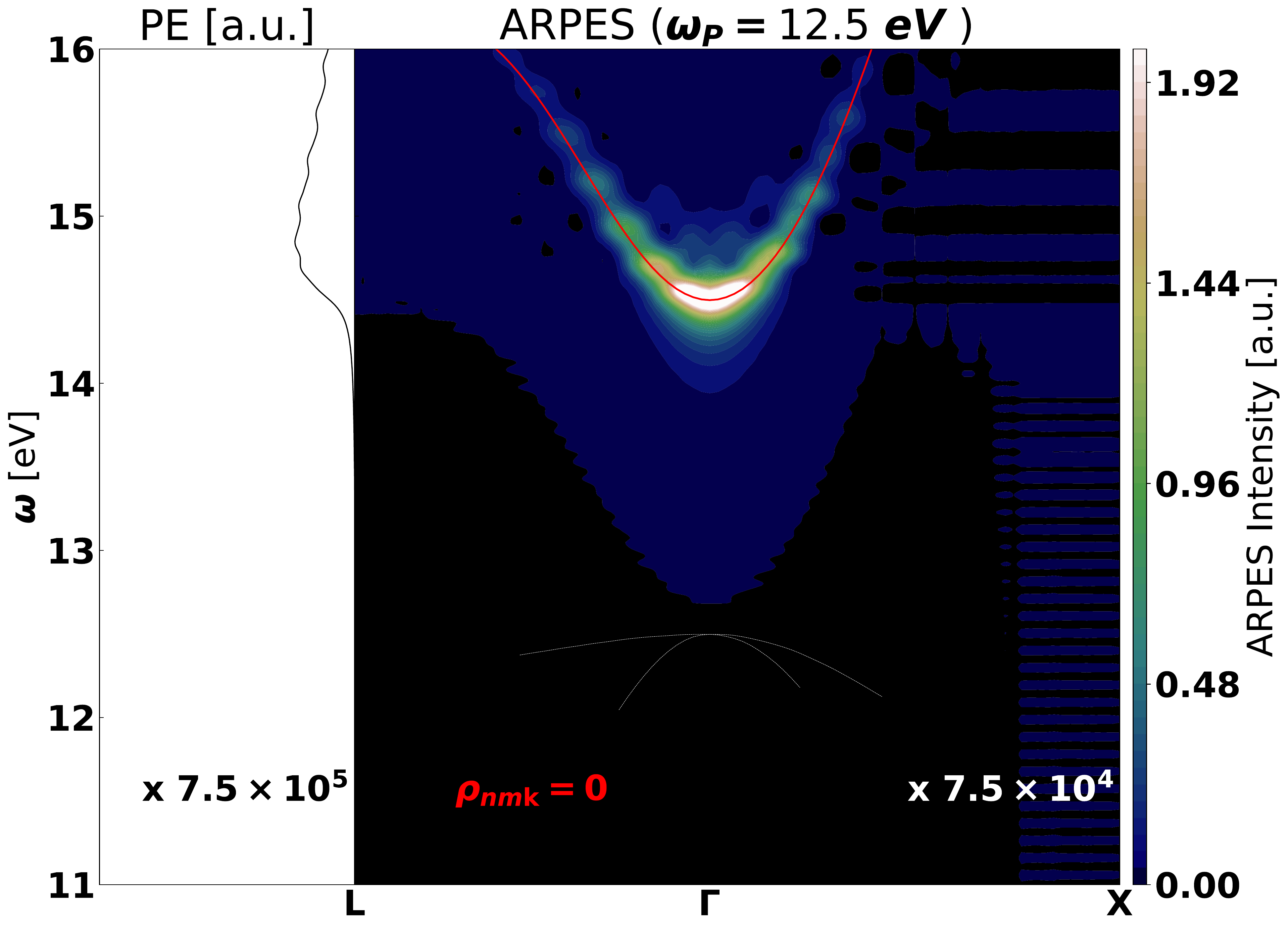}
	\end{center}
\caption{TR-ARPES spectral function, and corresponding $\blk$--integrated photo--emission (PE) signal of LiF. (left panel) Signal from carriers generated with a laser pulse with $\omega_P=14.8$~eV; coherent (a) and non-coherent (c.a) carriers give exactly the same signal. (central panel) Signal from coherent excitons (b) generated with a laser pulse with $\omega_P=12.5$~eV. (right panel) Signal from the carriers distribution reached, starting from the coherent exciton state, after sending to zero the off-diagonal elements of the density matrix (c.b). A replica of the valence band at energy $\epsilon_{v\blk}+\omega_P$ is shown in white. The numbers reported in the PE and ARPES panels represent the magnification factor needed for the signal in the conduction band region to reach similar intensity as the signal from the valence band (see also Supplemental Material).}
\label{fig:TR-ARPES}
\end{figure*}
We first analyze the TR-ARPES signal, shown in fig.~\ref{fig:TR-ARPES} for of the four states generated. The spectral function is defined
from the time propagation without probe pulse as $I(\blk,\omega)=-i\sum_{cc'} G^{<}_{cc'\blk}(\omega)$, obtained via the Fourier transform of the GKBA Green
function~\footnote{This definition of the TR-ARPES signal implicitly assumes that the time resolution is longer than the fast oscillations at the frequency of the exciton, and accordingly the terms $G^{<}_{cv\blk}(t,t')$ can be neglected. This is a necessary condition if one wants to keep energy resolution in the spectral signal. If the probe pulse is short enough then $\{t,t'\}\rightarrow t_{probe}$ and one is left with an energy integrated signal which oscillates in time as discussed in Ref.~\onlinecite{Perfetto2019b,Rustagi2019}.}
\begin{equation}
G^{<}_{cc'\blk}(t,t')= \sum_{n} \rho_{cn\blk}(t) G^{(r)}_{nc'\blk}(t,t')- G^{(a)}_{cn\blk}(t,t') \rho_{nc'\blk}(t') .
\label{eq:GKBA}
\end{equation}
The TR-ARPES spectral function generated from free carriers, after the pump at $\omega_{P_a}=14.8$~eV, shows a finite signal on the conduction band of LiF (see fig.~\ref{fig:TR-ARPES}, left panel). The same numerical result can be obtained at the TD-IP and at the TD-HSEX level in this case (see Supplemental Material).
At the IP level the retarded and advanced propagators remains frozen at the equilibrium value and are diagonal: ${G^{(r/a)}_{nm\blk}(t,t')= \delta_{n,m} e^{-i\epsilon_{n\blk}(t-t')}}$. As a consequence only the terms $\rho_{cc'\blk}$ enter eq.~\eqref{eq:GKBA}. Since we have here just one conduction band well separated from the others, the ARPES signal from the coherent carriers state (a) and the dephased carriers (a.c) are in practice identical. 
In the figure it is also reported a replica of the valence band shifted up by 14.8~eV (i.e. the pump frequency). The ARPES signal is localized where the VB replica crosses the CB thus, for the $\{v_{P_a},c_{P_a},\blk_{P_a}\}$ which respect the energy conservation at the IP level $\omega_P=\epsilon_{c\blk}-\epsilon_{v\blk}$.
The same signal can be obtained expressing the TR-ARPES in terms of the occupations only $I(\blk,\omega)=\sum_{c} f_{c\blk}\delta(\omega- \epsilon_{c\blk})$. This result can be easily obtained from Eq.~\eqref{eq:GKBA} using the IP expression for the retarded and advanced propagators and the definition of the electronic populations in terms of the diagonal elements of the density matrix.
In presence of nearly degenerate conduction bands (i.e. when $\Delta\epsilon_{cc'\blk}\leq 2\pi/\sigma$, $\rho_{c\neq c'\blk}$ terms may be activated and have important implication in specific cases. As an example $\rho_{c\neq c'\blk}$ terms have key importance in defining the spin-resolved (SR) TR-ARPES signal of free carriers injected via optical orientation experiments in GaAs~\cite{Dalessandro2020}.

The TR-ARPES spectral function generated from the coherent exciton state (b), after the pump at ${\omega_{P_a}=12.5}$~eV, is instead a replica of the valence band shifted by the exciton energy and weighted by the exciton wave-function (see fig.~\ref{fig:TR-ARPES}, central panel).
In this configuration the TD-HSEX retarded and advanced propagators, ${G^{(r/a)}_{nm\blk}(t,t')}$ reconstructed from the time--ordered exponential of the HSEX hamiltonian, acquire non diagonal terms which, multiplied by the fast oscillating $\rho_{nm\blk}(t)$ terms, determine the TR-ARPES signal. For the coherent exciton state generated by the pump pulse, in the low--density regime, an exact expression can be derived for $G^{<}(t,t')$ exploiting the mapping between the state generated by the pump pulse and the state generated via the Matsubara procedure discussed in Ref.~\onlinecite{Perfetto2019b}: 
\begin{equation}
{G^{<}_{cc'\blk}(\omega)=(2\pi i) N_c \sum_v A^{\lambda_1\mathbf{0}}_{cv\blk}A^{\lambda_1\mathbf{0},*}_{c'v\blk} \delta\left(\omega-(\epsilon_{v\blk-\blq}+\omega_{\lambda\blq})\right)}
\label{eq:Gless_exc}
\end{equation}
Such expression perfectly describes the signal obtained via the real--time propagation and shown in Fig.~\ref{fig:TR-ARPES}, central panel. The state corresponds to an exciton population peaked at the lowest energy exciton $\{\lambda_1\mathbf{0}\}$, i.e. ${N_{\blq\lambda}=N_{exc}\delta_{\lambda\lambda_1}\delta(\blq)}$, with $E_{\lambda_1\mathbf{0}}=12.5$~eV. Due to its coherent nature, it is referred to as nonequilibrium exciton BEC. The prefactor determining the intensity of the signal corresponds to the mean~\footnote{A coherent state is not an eigenstate of the population operator, thus we can only speak about the mean excitonic population} value of the excitonic population ${N_{exc}=N_c}$~\cite{Moskalenko2000,Perfetto2019b}.
The TR-ARPES spectral function generated from the non--coherent state (c.b), after sending to zero the off-diagonal elements of the density matrix, is reported in fig.~\ref{fig:TR-ARPES}, right panel, and it is \emph{completely different} from the one of the central panel which also includes the off-diagonal terms of the density--matrix.
After the dephasing, the signal has contribution only from the conduction band, while its distribution in k-space is reminiscent from the excitonic wave-function $A^{\lambda_1\mathbf{0}}_{cv\blk}$. Indeed it can be shown that ${f_{c\blk}=\rho_{cc\blk}\propto \sum_v |A^{\lambda_1\mathbf{0}}_{cv\blk}|^2}$. Such drastic change proves the unphysical nature of the chosen decoherence procedure, which completely alters the state generated by the pump pulse.

As we already observed, it is non--trivial to define a proper dephasing process within the EOM for the one-body density matrix $\rho$ which leads to an exciton population (Fig.~\ref{fig:neq_states_overview}.d). The reason is that the exciton is a two--particles object and approximations to its description can be more easily obtained starting from the two-body density-matrix~\cite{Sangalli2018a}. Nevertheless, the
ARPES signal predicted from a non-coherent excitonic populations has been discussed in the literature. The signal has the form~\cite{Rustagi2018}:
\begin{equation}
{I[N_{\l\blq}](\blk,\omega)=2\pi \sum_{\l\blq cv}N_{\l\blq} |A^{\l\blq}_{cv\blk}|^2 \delta\left(\omega-(\epsilon_{v\blk-\blq}+\omega_{\l\blq})\right)}. 
\end{equation}
In the limit of a peaked excitonic distribution ${N_{\blq\lambda}\approx N_{exc} \delta_{\lambda\lambda_1}\delta(\blq)}$ again a replica of the valence band weighted by the excitonic wave--function is obtained.
Thus, regardless of originating from a coherent or a non-coherent state, the excitonic ARPES signal has the same shape~\cite{Christiansen2019,Perfetto2019b}. Recently this feature was exploited to try to measure experimentally the excitonic wave-function~\cite{Dong2020}. As at the IP level, one difference is that in the non--coherent case the terms with $c\neq c'$ are not accounted for.
Other differences could be found probing the system with a time resolution short enough to resolve the ultra-fast oscillations of the coherent excitonic state~\cite{Perfetto2020,Rustagi2019,Note2}. This however would in general lead to loss of energy resolution. 
The main message of this section is that the ARPES signal of a system with a finite excitonic population has a clear signature (Fig.~\ref{fig:TR-ARPES}.(b) and that such signal can be constructed fully \ai\, within TD-HSEX scheme.

\section{Transient Absorption}

We now turn to the analysis of the Tr-Abs signal from the four states discussed so far.  To this end I compute 
$\blP^{(i)}(t)=\blP_{Pp}(t)-\blP_{P}(t)$ obtained from the two time propagations with and without the probe. Again the superscript is a label for the corresponding state: $i=a,c.a,b,c.b$. The probe pulse used is a delta--like function in time which excites the whole spectrum. A dephasing term is applied when the Fourier transform of the polarization is performed to smooth the final result:
\begin{equation}
P^{(i)}(\omega)=\int_{t_p} dt\, \blP^{(a/c.a)}(t) e^{i\omega t - \eta (t-t_p)}.
\end{equation}
The transient signal is obtained subtracting the equilibrium contribution: $P^{(i)}(\omega) - P^{(eq)}(\omega)$. Following the derivation of Ref.~\onlinecite{Perfetto2015}, it can be shown that this can be approximated via an adiabatic response function constructed from the non equilibrium density matrix $\chi[\rho_{nm\blk}(\tau)](\omega)$.

\begin{figure}[t]
	\begin{center}
		\includegraphics[width=0.450\textwidth]{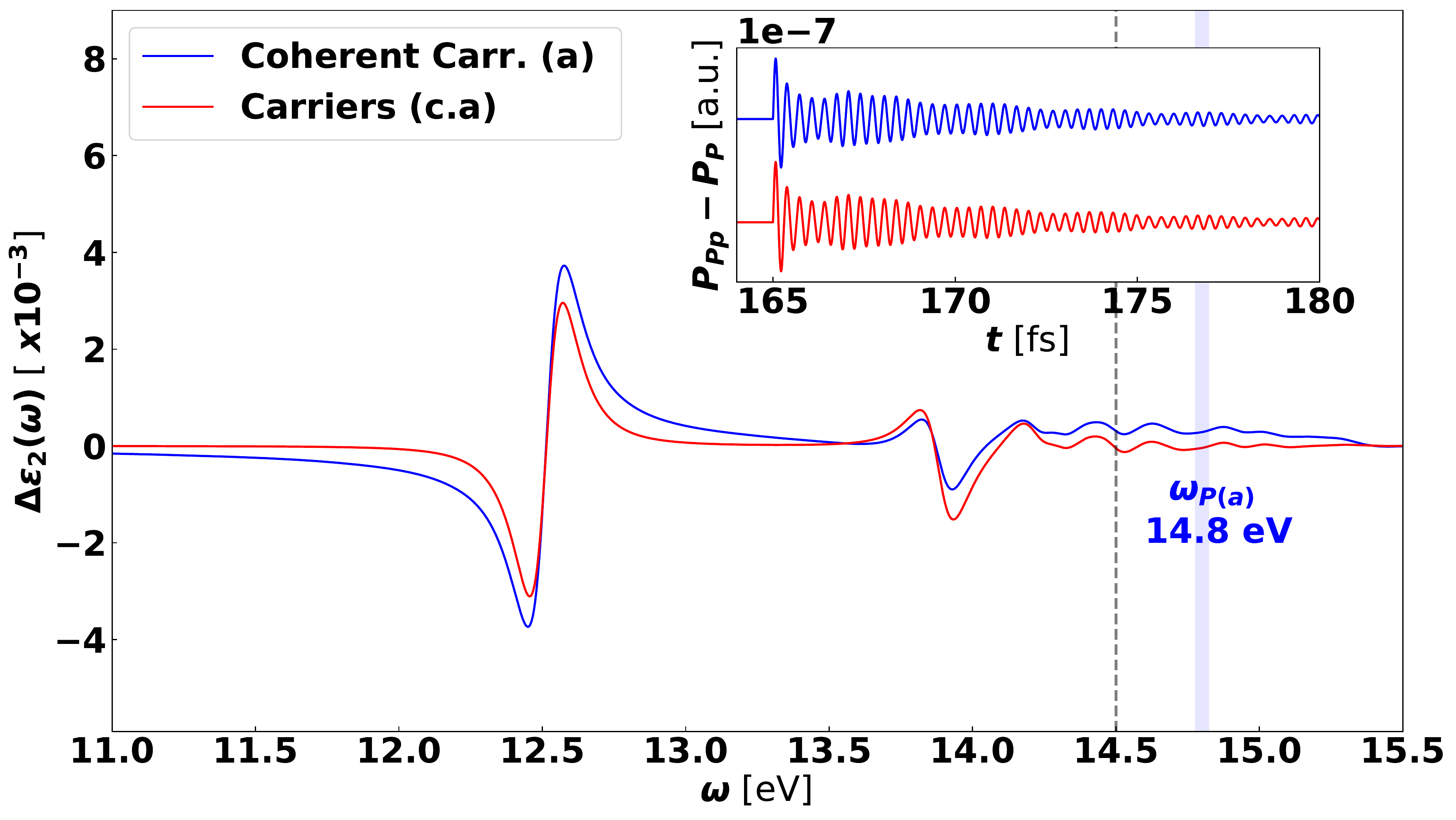}
		\includegraphics[width=0.450\textwidth]{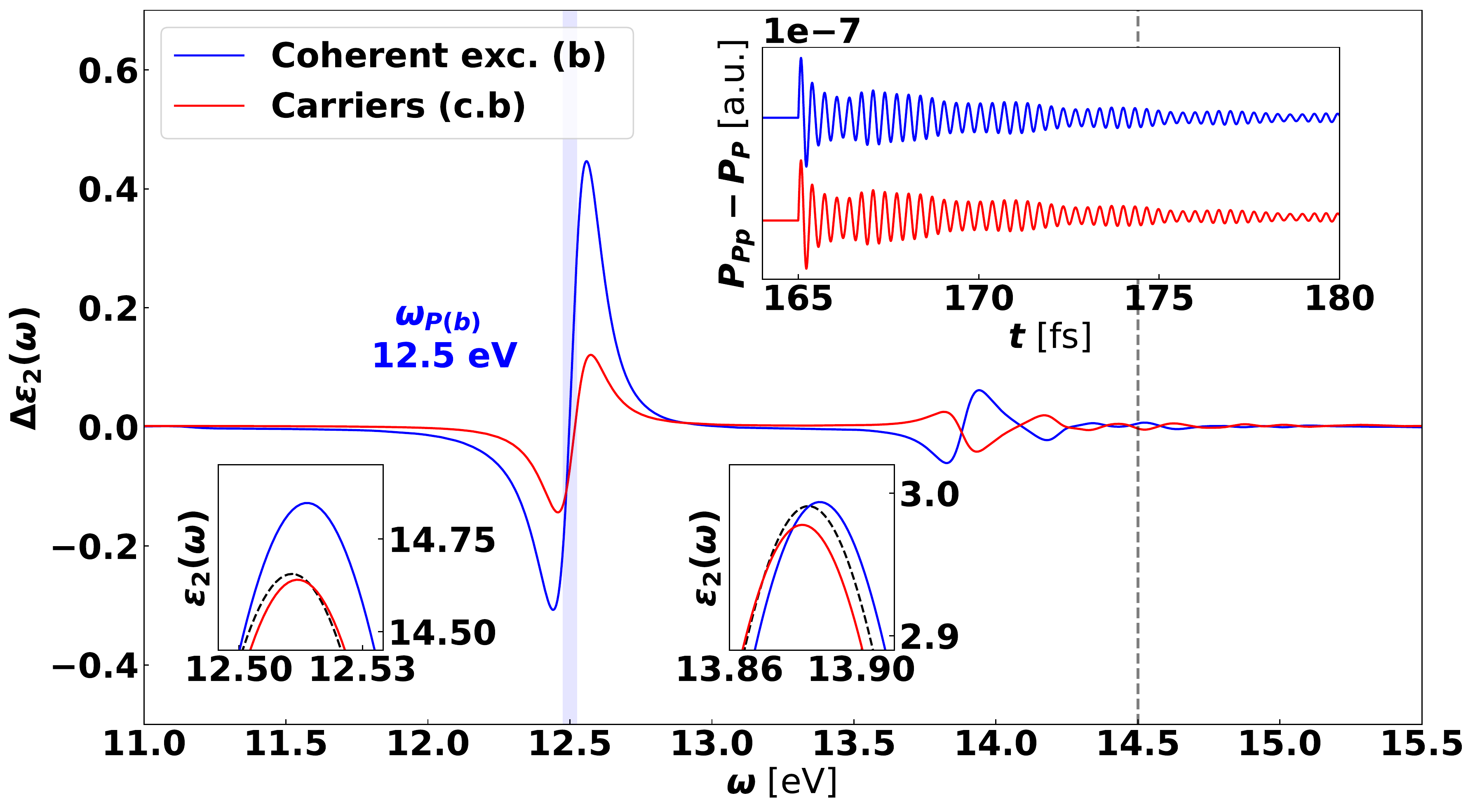}
	\end{center}
	\caption{Transient Absorption $\Delta\varepsilon^{(i)}_2(\omega)=\varepsilon^{(i)}_2(\omega)-\varepsilon^{(eq)}_2(\omega)$ for LiF shown for the four states introduced in Fig~\ref{fig:neq_states_overview}: coherent carriers (a) and non-coherent carriers (c.a), and coherent exciton (b) and non-coherent carriers distribution obtained dephasing the excitonic state (c.b). The probe induced polarization $\blP^{(i)}(t)=\blP_{Pp}(t)-\blP_{P}(t)$ is shown in the inset.
		The blue shaded line marks the laser pump frequency, the black dashed line the position of the quasi-particle gap. The two extra insets in the lower panel show a zoom of $\varepsilon^{(eq)}_2(\omega)$ (black line) and $\varepsilon^{(i)}_2(\omega)$ around the two main excitonic peaks.}
	\label{fig:TR-ABS}
\end{figure}

In the top panel of fig.~\ref{fig:TR-ABS}, the Tr-Abs generated from the coherent carriers state (a) and non-coherent carriers state (c.a) are compared. The inset shows a plot of the corresponding $P^{(a/c.a)}(\omega)$ polarization.
While the pump step can be described at the IP level for the pump resonant with the continuum, in the probing step excitonic effects cannot be neglected. This is why, to construct the Tr-Abs signal, I start from the state generated at the IP level, but I further propagate the EOM in the time window 160-260 fs at the HSEX level. A detailed discussion on this point can be found in the Supplemental Material. Since the EOM is propagated at the HSEX level, the signal corresponds to the adiabatic response function
obtained solving the Bethe-Salpeter equation $\chi^{BSE}[\rho_{nm\blk}(\tau)](\omega)$. The adiabatic approximation gives the exact result if the time duration of the probe process is much slower than the dynamics of the density matrix. This is the case in the non--coherent carriers state (c.a), where the density matrix is time independent.
The adiabatic formulation offers a clear interpretation of the transient signal: due to the non equilibrium density matrix, on one side there is a reduction or bleaching of the transition due to Fermi blocking, on the other side both the QP energies and the excitonic energies are renormalized~\cite{Sangalli2016}. This explains the overall derivative signal, which is dominated by the energies renormalization, especially in the bound region.
In the coherent carriers state (a), corrections due to the time oscillating terms $\rho_{n\neq m\blk}(t)$ exist (as opposed to the TR-ARPES signal of the system driven in the continuum). However, they give here a negligible contribution as shown in fig.~\ref{fig:TR-ABS}.

In the bottom panel of fig.~\ref{fig:TR-ABS} the Tr-Abs generated from the coherent exciton state (b) and the non--coherent state (c.b) are compared. 
The inset shows a plot of the corresponding $P^{(b/c.b)}(\omega)$ polarization.
While in the non--coherent case the signal can be interpreted via the adiabatic response function $\chi^{BSE}[\rho_{nm\blk}(\tau)](\omega)$,
in the coherent case (b) the adiabatic approximation is not justified, because there is a dominant fast oscillation in $\rho_{n\neq m\blk}(t)$.
To analyze the transient absorption signal we can inspect the EOM for $\rho_{cv\blk}$, eq.~\eqref{eq:EOM_rho}. 
The term involving the variation of the HSEX self-energy can be expanded to linear order in the probe, i.e. I take either the self-energy, $\Delta\Sigma^{Hxc,P}$, or the density matrix, $\rho^P_{cv\blk}$, evaluated without the presence of the probe. As discussed in the Supplemental Material, these two terms give rise to a renormalization of the excitonic peaks. Finally, the term $[U^{ext},\rho]_{cv\blk}$ can be seen as the source of potential bleaching of anyway changes in the intensity of the peaks.
This explains (without the need of the adiabatic approximation) the derivative signal at the excitonic energies shown in fig.~\ref{fig:TR-ABS}.
Both the for coherent excitonic state (b) and the non--coherent carriers state (c.b)  the signal is mostly localized nearby the pumping energy $\omega_{P_b}=12.5$~eV, due to a blue shift of the excitonic peak at 12.5~eV.
The similarity between the signal of the two states (b) and (c.b) is in sharp contrast with the TR-ARPES case, where the dephasing procedure completely changed the nature of the signal. The reason is twofold. In TR-ARPES (i) the electron--hole interaction has a negligible role in the interaction of the probe pulse with the system since the probe does not generate excitons. However (ii) the signal is directly affected by the excitons generated by the pump pulse since the exciton binding energy must be supplied to break them and extract carriers. On the other hand, in the Tr-Abs case (i) the probe generates coherent excitonic states in the energy window of 12.5~eV and the electron-hole interaction has a key role in this process, and (ii) the signal is only indirectly affected by the presence of the excitons generated by the pump pulse.
Nevertheless, some differences between the carriers and the coherent exciton case remain. At the lowest energy excitonic peak, at 12.5~eV, the peak intensity is reduced for the non--coherent state (c.b), while it is enhanced for the coherent excitonic state (b). The effect on the peak around 14.0~eV is even more striking with an opposite shift in the two cases.

We now try to define the expected signal for the non--coherent excitonic state (d), starting from the excitonic operators, ${\hat{e}^{\dag}_{\lambda\blq},\hat{e}_{\lambda\blq}}$, and use a non-interacting boson approximation, i.e. assuming that (i) the excitonic operator follows the bosonic commutation rules and that (ii) the energy needed to generate an exciton is independent of the exciton population. Doing so I also neglect terms where the probe promotes an existing exciton to a higher energy state. Such transition ${(\w_{\l\blq}-\w_{\l'\blq})}$ require to consider the internal structure of the exciton, and are expected to be important in the low energy range of the spectrum~\cite{Koch2006}.
The BSE response function can be defined starting from the (retarded) excitonic propagator
\begin{equation}
L_{\l\l'}^{BSE}(\w)=\d_{\l\l'}
\left(\frac{1+N_{\l\mathbf{0}}}{\w-\w_{\l\mathbf{0}}+i\eta}-  
\frac{N_{\l\mathbf{0}}}{\w-\w_{\l\mathbf{0}}+i\eta}\right) , 
\end{equation}
that is independent on the excitonic population.
Indeed the emission term, proportional to $N_{\l\mathbf{0}}$ is exactly compensated by the enhancement of the absorption term due to the factor $1+N_{\l\mathbf{0}}$. We mention that, on the contrary, time-resolved photoluminescence (TR-PL) experiments can be well described starting from the bosonic approximation to the exciton~\cite{Moskalenko2000} since the emission terms only is needed.
The BSE response function $\chi^{BSE}(\omega)$ constructed from $L_{\l\l'}^{BSE}(\w)$ is identical to the equilibrium one.
This means that, while in TR-ARPES the signal can be modelled, as a first approximation, starting from the bosonic nature of the exciton, in Tr-Abs the same approximation gives zero signal, and the corrections to the bosonic nature of the exciton need to be considered.
This is why in Fig.~\ref{fig:neq_states_overview}.d there is not an expression for $\chi^{BSE}[N_{\lambda\blq}](\omega)$.  Nevetheless the above analysis gives us some hints. The absence of bleaching is in contrast with what would happen in presence of free carriers populations, where both the emission term (proportional to $f_{c\blk}(1-f_{v\blk})$) and the absorption term (proportional to $f_{v\blk}(1-f_{c\blk})$) are bleached due to Pauli Blocking. Indeed, as we observed looking at the numerical results, the excitonic peaks are bleached in presence of carriers, while they are enhanced in presence of coherent excitons. The results presented in the present manuscript show that also in Tr-Abs experiments the signal due to carriers and excitons can be distinguished, although the effect is less striking than the TR-ARPES case. A more detailed analysis is needed to further clarify this point, including a scheme able to compute numerically the Tr-Abs from a non--coherent excitonic state. 

\section{Conclusion}

In conclusion, I presented a scheme to describe pump and probe experiments including the physics of the exciton. The approach has been implemented in the real--time module of the yambo code~\cite{Sangalli2019}, and tested both in the case where LiF is driven in resonance with the excitonic state and in the case where the pump energy is in the continuum. The results highlight the importance of excitonic effects first of all in the TR-ARPES case, but also in the description of Tr-Abs experiments.
Part of the discussion has been focused on the transition from a coherent state to a non--coherent population. 
Recently published results, based on the interpretation of pump and probe experiments via \ai\, simulations that do not fully account for excitonic effects, could be reviewed~\cite{Pogna2016,Smejkal2021,Wang2018,MolinaSanchez2017}. The approach can become the reference scheme to describe the generation and the detection of coherent excitons in a wide range of materials, in the same way, the \ai\, Bethe--Salpeter equation become the reference scheme for modelling excitons at equilibrium.
Different extensions are possible, including the update of the screening to capture the exciton Mott transition at higher pump fluences or the inclusion of the interaction with phonons and photons to properly model the dynamics of the coherent state, exciton dephasing processes, and exciton lifetimes~\cite{Palummo2015,Paleari2019,Cudazzo2020,Chen2020}.

\subsection*{Acknowledgments}

I acknowledge funding from MIUR (Italy), PRIN Grant No. 20173B72NB, from the European Union, project MaX Materials design at the eXascale H2020-EINFRA-2015-1, (Grants Agreement No. 824143), and project Nanoscience Foundries and Fine Analysis-Europe H2020-INFRAIA-2014-2015 (Grant Agreement No. 654360).

\bibliography{manuscript}

\end{document}


\title{Supplemental Material. Excitons and carriers in transient absorption and time--resolved ARPES spectroscopy: an abinitio approach}

\author{D. Sangalli}
\affiliation{Istituto di Struttura della Materia of the National Research 
	Council, Via Salaria Km 29.3, I-00016 Montelibretti, Italy}

\maketitle

\section{Convergence for the absorption spectrum and the excitonic wave--functions} \label{abs:convergence}
\subsection{Computational details for equilibrium properties and band structure}
The ground state properties of LiF are computed with an energy cutoff on the wave--functions of 80~Ry and a k-points grid $6\times 6\times 6$. Then a non-self consistent calculation is performed on three k-points grids, i.e 16x16x16, 24x24x24, and 32x32x32 with 100 bands.
\begin{figure}[h]
	\includegraphics[width=0.33\textwidth]{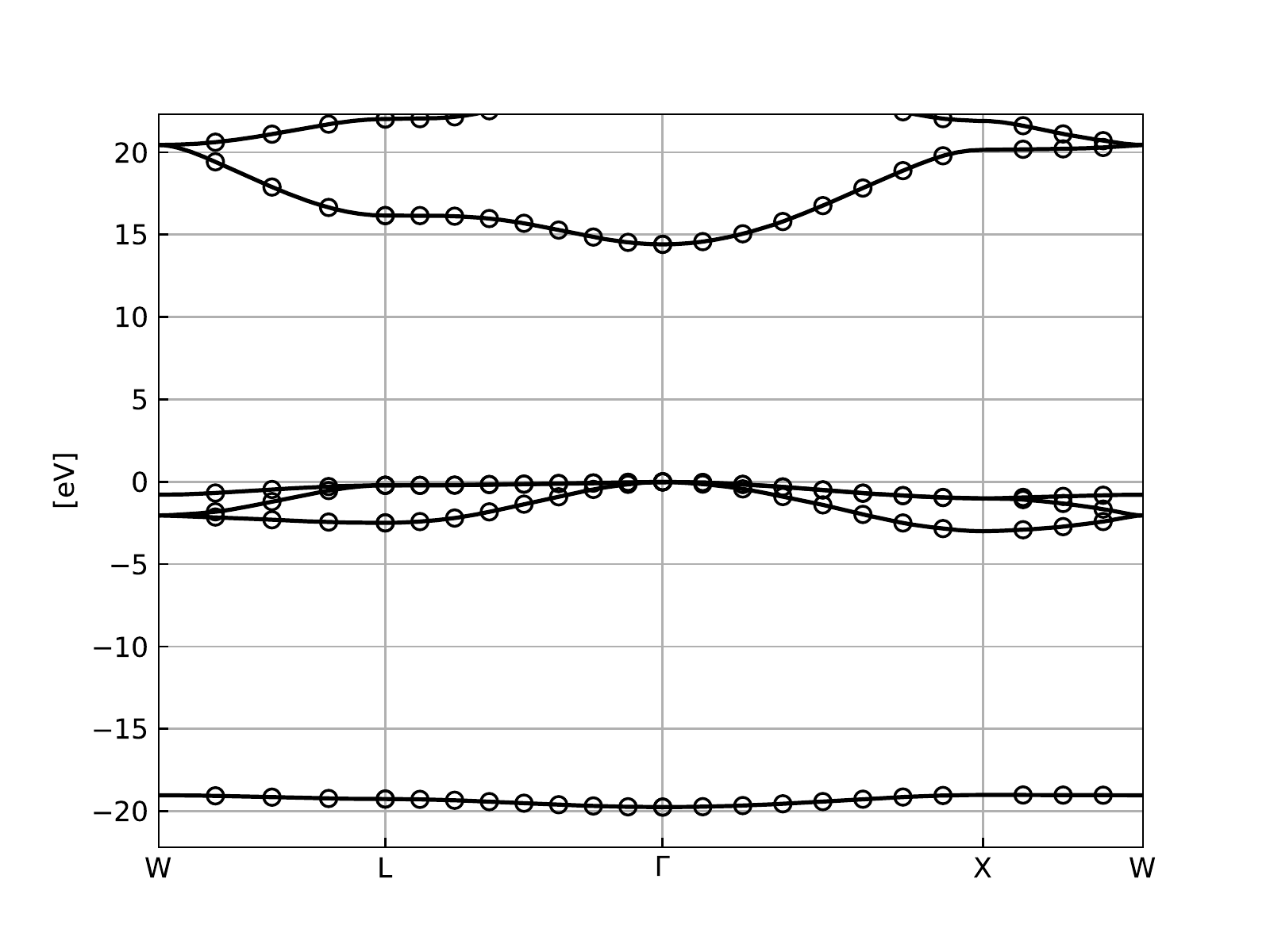}
	\includegraphics[width=0.33\textwidth]{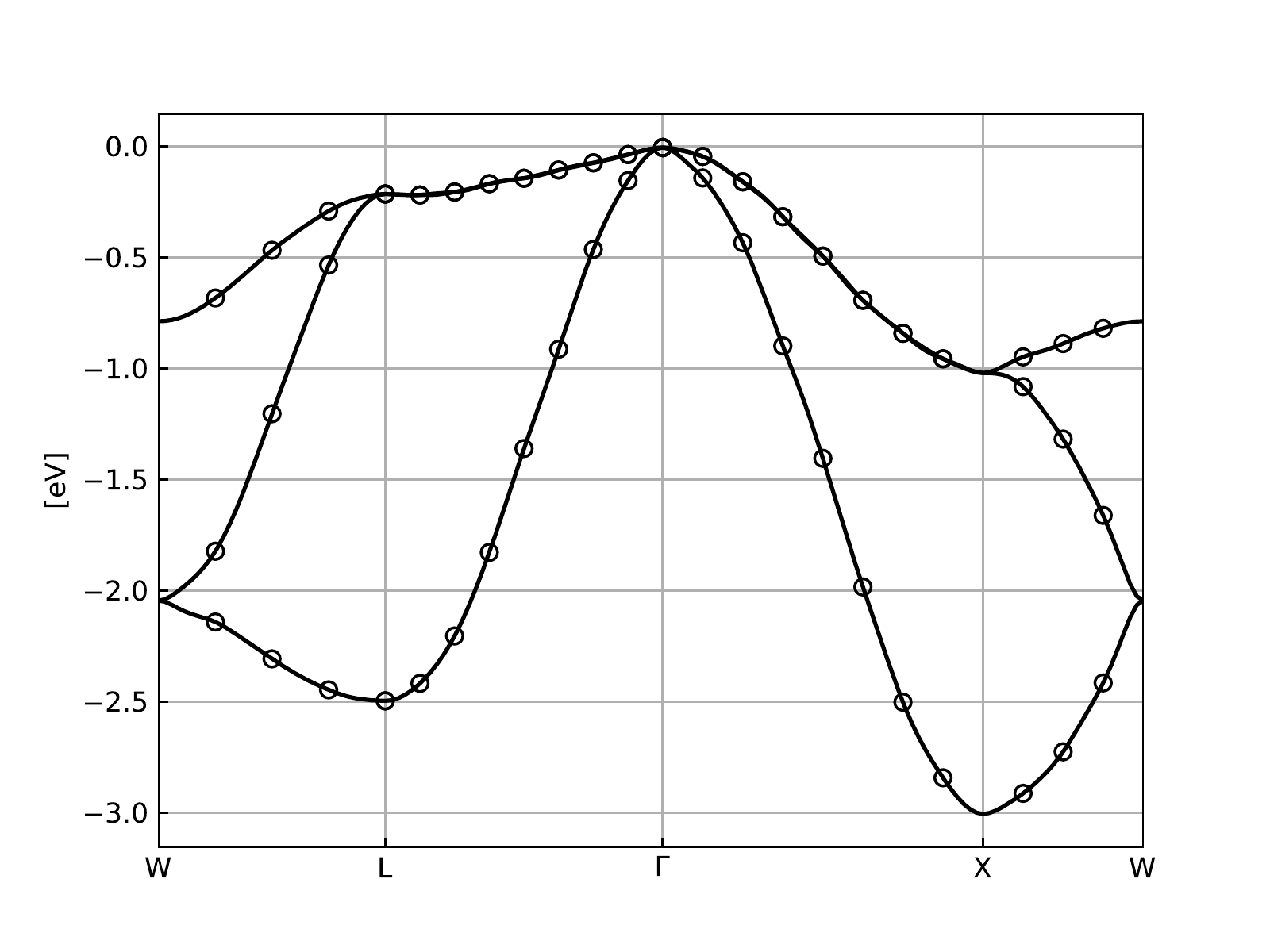}
	\includegraphics[width=0.33\textwidth]{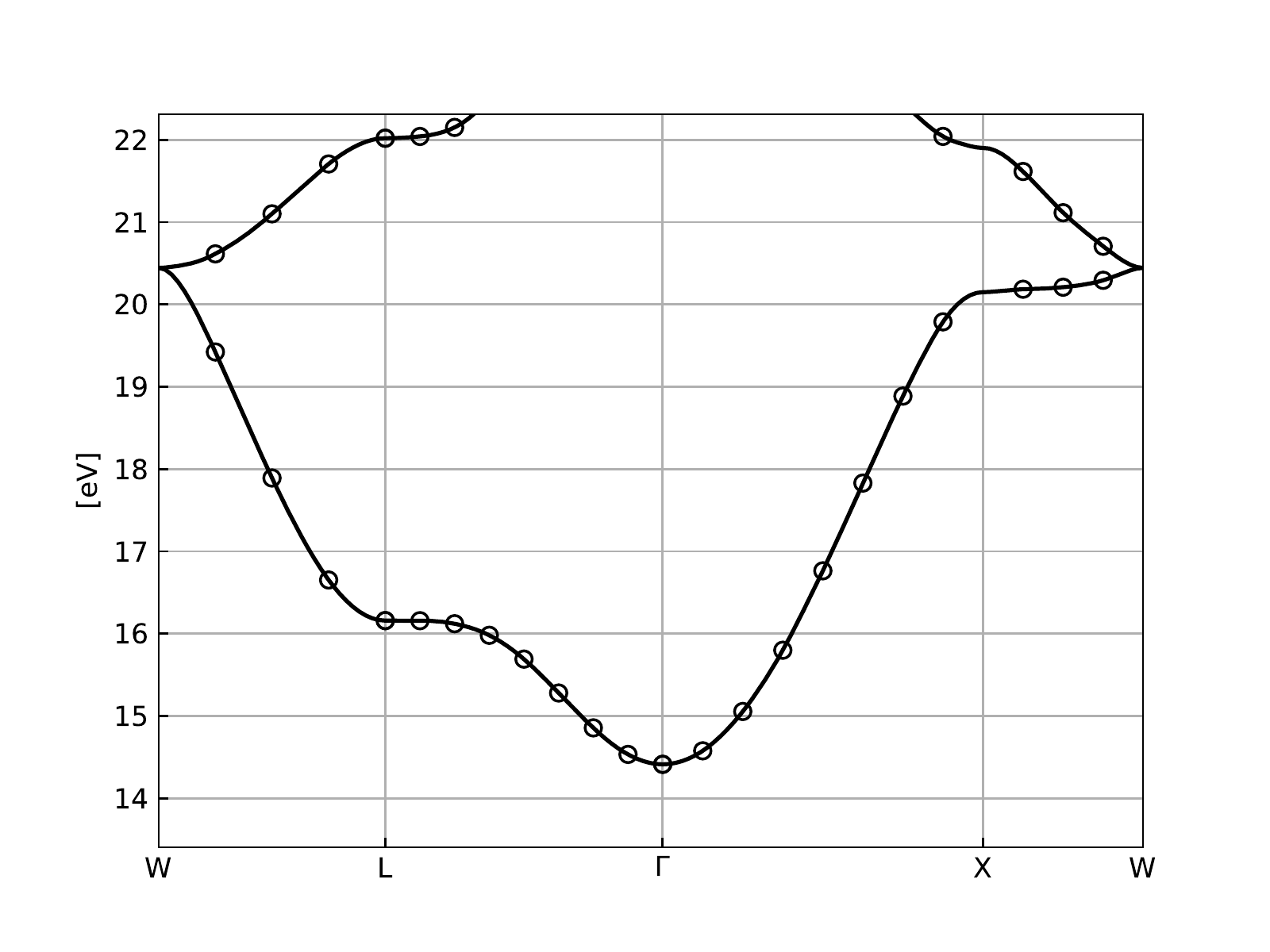}
	\caption{(left panel) Band structure in the energy range $-21$ - $21$~eV which contains the first 5 bands of LiF. (central panel) zoom in the valence band region, bands from 2 to 4. (right panel) zoom in the conduction band region, band 5. Points correspond to calculated eigenvalues on the 16x16x16 grid, while lines are obtained via 3D Fourier interpolation.} 
	\label{fig:Bands}
\end{figure}
In Fig.~\ref{fig:Bands} the band structure obtained from a Fourier interpolation starting from the 16x16x16 is shown. The first occupied valence band lies at $\approx -20$~eV, then there are three which are degenerate at $\Gamma$ with $\epsilon_{v\Gamma}=0.$~eV and span an energy range of $\approx 3$~eV. The first conduction band determines the gap with $\epsilon_{c1\Gamma}=14.5$~eV spans an energy range of $\approx 6$~eV.
A shissor operator of 6.05~eV is applied in the plot to shift the KS band gap. The second conduction band start at $\epsilon_{c2\Gamma}=20.0$~eV.

\subsection{Absorption spectrum}

For the calculation of the BSE matrix, bands from 2 to 5 are then considered (for the 16x16x16 we also verified that using bands from 1 to 6 no significant changes are observed). Indeed the critical parameter for the convergence of the spectrum is the sampling of the Brillouin zone, as discussed more in details in the present section. 
The other parameters used for the BSE calculations are: 965 Reciprocal lattice vectors for the exchange, 59 reciprocal lattice vectors and 100 bands for the screening, 59 reciprocal lattice vectors for the screened electron--hole interaction. The BSE is then solved using the Haydock solver to obtain the spectrum and the SLEPC library to obtaine the first 500 poles of the BSE matrix.

Here I report the convergence of the equilibrium absorption spectra and amplitudes considering a $\blk$--grid up to $32\times 32\times 32$. Solving the BSE on such grid is demanding but feasible (there are $32^3\cdot3=98.304$ transition, and the BSE has 9.663.676.416 matrix elements).
We are not able to perform real--time simulations on the same grid, mostly because the real-time code is not as much optimized as the BSE code.
\begin{figure}[h]
	\includegraphics[width=0.95\textwidth]{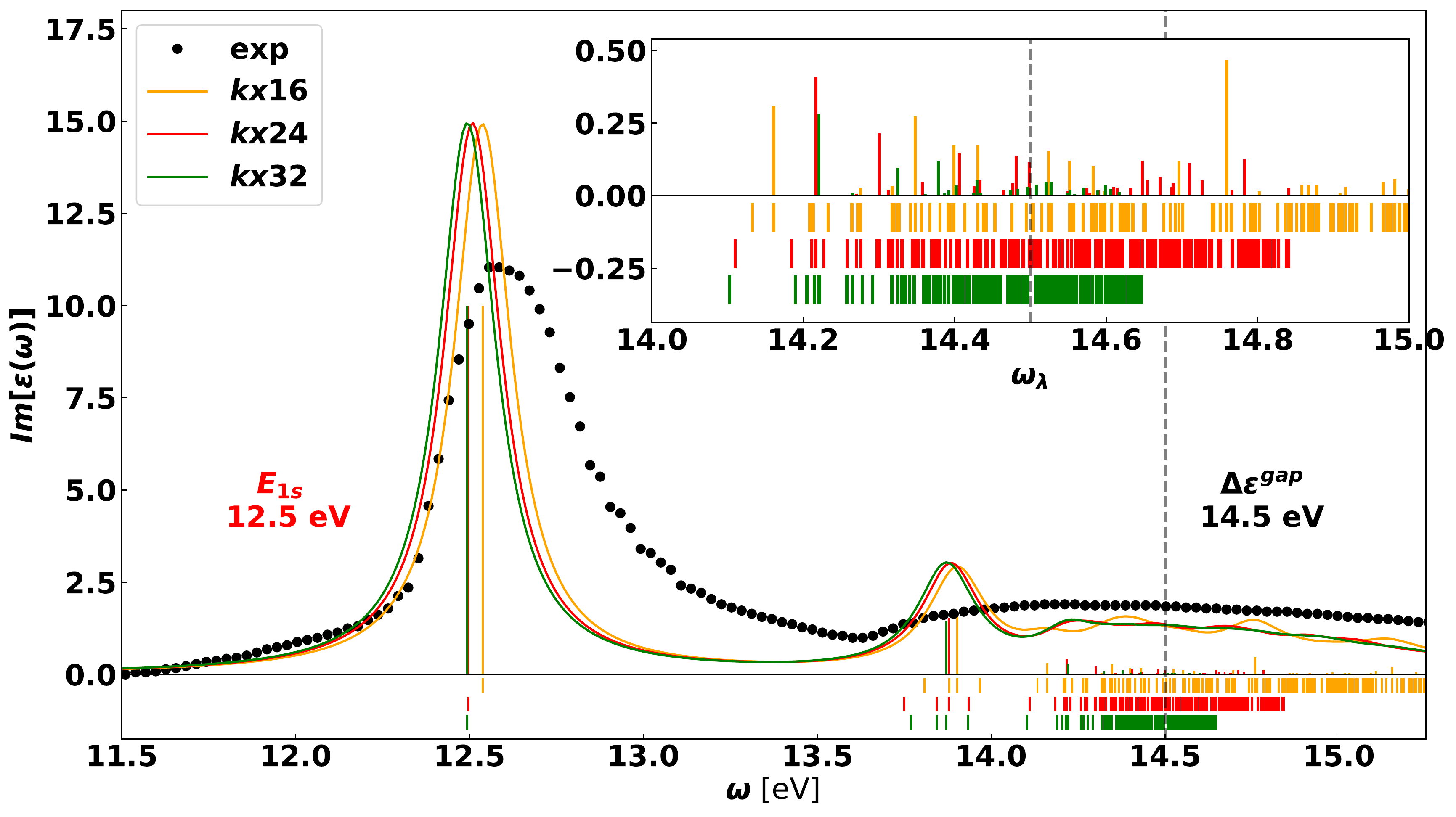}
	\caption{Convergence of the absorption spectrum with k-grids 16x16x16 (orange), 24x24x24 (red) and 32x32x32 (blue). For each simulation the first 500 eigenvalues of the excitonic hamiltonian are reported both weighted by the oscillator strengths (under the spectrum, positive $y$-axis) and as a list of bars also including dark transitions ()negative $y$-axis).} 
	\label{fig:ABS_convergence}
\end{figure}
The spectrum is fully converged with the $24\times 24\times 24$ grid. Already the $16\times 16\times 16$ grid gives a very good result, with the energy of the first two bright excitonic peaks within 50~meV from the denser grid and few oscillations in the continuum. In the continuum convergence is achieved using a smearing paramenter $\eta=0.1$~eV.
This means that real-time simulations would be converged in such energy range using the same smearing parameter.
In the limit of very small smearing parameter we need to look at the BSE eigenvalues. At the BSE level these are reasonably converged with the $24\times 24\times 24$ grid in the bound region of the spectrum, but still not converged in the continuum. For example the region around $E=14.8$~eV which we consider in the main text has a very dense spectrum of poles. For the $24\times 24\times 24$ we obtain 79 poles in a region of 50~meV (14.775~eV$<E<$14.824~eV), i.e. an average of 3 poles each 2~meV. However only very few poles have non negligible intensity in the absorption spectrum (see inset of fig.~\ref{fig:ABS_convergence}). In the limit of infinite  k-points sampling we expect the number of poles to become infinite in order to capture the absorption in the continuum. However it is not possible to reach such situation numerically. 

\subsection{Excitonic wave-function}
\begin{figure}[t]
	\includegraphics[width=0.45\textwidth]{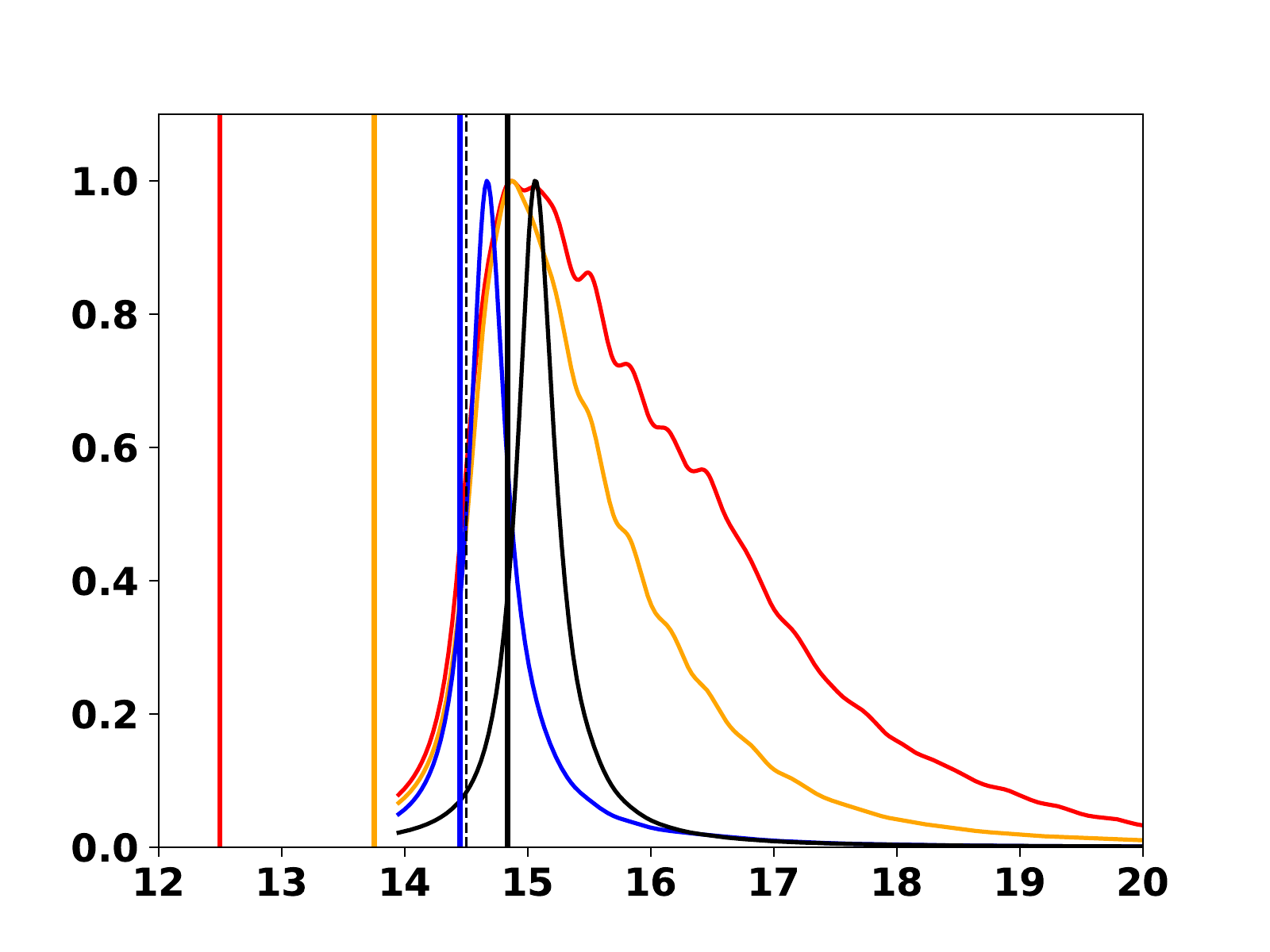}
	\includegraphics[width=0.45\textwidth]{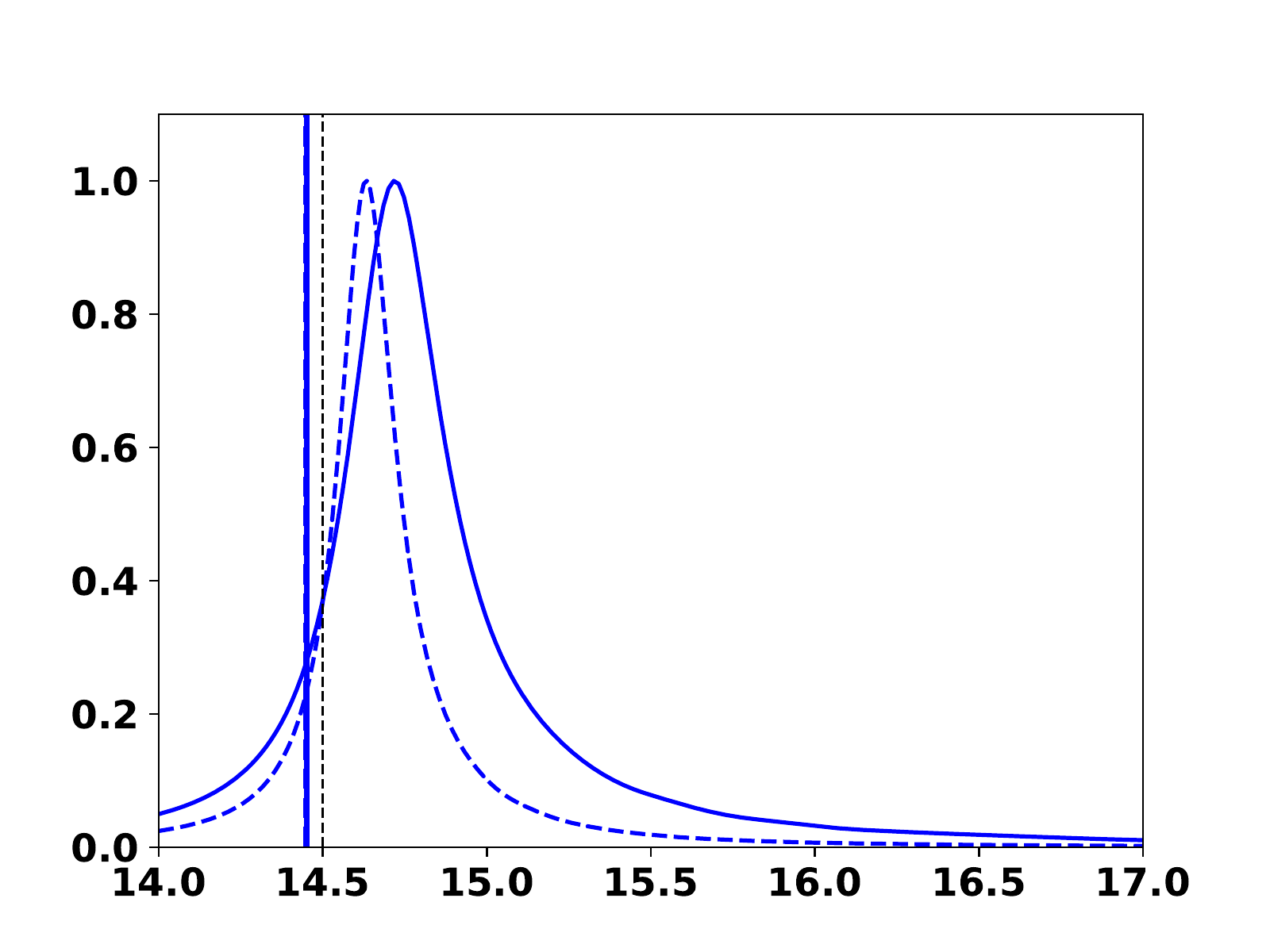}v
	\caption{(left panel) Representation of the excitonic amplitudes $F^{\lambda}_{\eta}(\omega)$ defined in eq.~\eqref{eq:exc_amplitude} for four different 
	states and $\eta=0.1\ eV$, with the k-grid 24x24x24. The matching energy $E_\lambda$ is represented as a vertical bar with the same color for each state. The QP gap is represented by a black dashed line. (right panel) the excitonic amplitude just below the QP gap is compared for two different k-grids: 24x24x24 (blue continuous) and 32x32x32 (blue dashed). The result slowly tends to a delta like function.} 
	\label{fig:exc_amplitudes}
\end{figure}

We now move to the analysis of the excitonic wave--functions $A^{\lambda}_{cv\blk}$. To this end I 
introduce the excitonic amplitudes which are a $\blk$-integrated representation of the excitonic wave--function:
\begin{equation}
F^{eh(\lambda)}_{\eta}(\omega)=\sum_{cv\blk}\frac{|A^{\lambda}_{cv\blk}|^2}{\omega-\Delta\epsilon_{cv\blk}+i\eta} .
\label{eq:exc_amplitude}
\end{equation}
%
In Fig.~\ref{fig:exc_amplitudes} we see that for the lowest energy exciton, $\lambda=1s$, $F^{eh(1s)}_{\eta}(\omega)$ has a very broad distribution and
involves electron--hole pairs up to more than $18\ eV$, i.e. $6\ eV$ above the $\omega_{1s}$.
This is due to the fact that the excitonic wave--function is quite localized in space and requires many elements in
the $\psi_{c\blk}\psi^*_{v\blk}$ basis set to be represented.
However there is no relation between the poles of $F^{eh(1s)}_{\eta}(\w)$ and the peak position $E_{1s}$ (which is marked by a vertical line in the figure).
The comparison between the pole $E_{1s}$ and the pole of $F^{eh(1s)}(\omega)$
is quite similar to the comparison of the (AR)PES signal of the coherent exciton state and the dephased state discussed in the main text.
Indeed the sole difference is how the energy of the valence band enters: $\Delta\epsilon_{cv\blk}$ and $\omega_\l$ vs $\epsilon_{c\blk}$ with $\omega_\l+\epsilon_{v\blk}$.

It is interesting to inspect higher energy states.
Above the QP gap, $F^\l_\eta(\w)$ are quite localized in energy.
This means that such $A^{\lambda}_{cv\blk}$ states are more ``delta like'' and correspond to
specific electron--hole pairs. In Fig.~\ref{fig:exc_amplitudes} the state with index 121,
at energy slightly below the QP gap, and the state with index 500, at energy above the QP gap are shown.
For such states the maxima of $F^{\lambda}_{\eta}(\omega)$ move in the same way as the energies $E_\lambda$. Compared to the maxima of $F^{\lambda}_{\eta}(\omega)$, $E_\lambda$ is just shifted at lower energy by a constant.
If the excitonic state is delta--like, the energy shift would be roughly due to the diagonal terms of the
eh--interaction entering the BSE for such specific transition: $W_{cv,cv}=\int d^3\blk W_{cv\blk,cv\blk} A^\l_{cv\blk}$.
In the limit of $A^\l_{cv\blk}\ra\delta(\blk-\blk_0)$ such energy shift should go to zero
since $W_{cv\blk,cv\blk} \approx 1/N_{\blk}$. However $A^\l_{cv\blk}\neq\delta(\blk-\blk_0)$ since
it must be ortogonal to $A^{1s}_{cv\blk}$. A converged quantitive value of such energy shift
as a function of the k-grid would however be very difficult to achieve and it is beyond the scope of the present work. The goal of the present discussion is to underline why in practice it would be very difficult to converge the TD-HSEX simulations with the laser pulse tuned in the continuum. As a consequence it is better to run TD-IP simulation in such case. In a following section, we further analyze such hypothesis, comparing the TR-ARPES spectrum of an HSEX simulations with the laser pulse in the continuum, with the IP result.

\clearpage

\section{Non-equilibrium states generated by the pump pulse}  \label{sec:neq_states}

\subsection{Computational details for the real-time simulations}
The same parameters used for BSE  (i.e. energy cutoff and number of bands) are also used for the real time propagation, with a time step of 5~as. $\epsilon_{thresh}=10^{-5}$~Hartree is chosen as the numerical tolerance for considering two energy levels $n$ and $m$ degenerate, so that $\rho_{nm}$ is not subject to dephasing (see main text). The same energy threshold is used for dipoles (see sec.~\ref{sssec:IP_approx} in the present SI). The real-time propagation has roughly the same computational cost of a BSE calculations. In particular the variation of the self energy can be constructed writing $\Delta\Sigma^{HSEX}=K^{HSEX}\,\Delta G^<$, with $K^{HSEX}$ the BSE kernel. There is, however, an advantage here. For BSE, $K^{HSEX}$ needs to be constructed in the full BZ, for real time propagation it needs to be constructed only for an `extended IBZ’, i.e. the IBZ defined by the symmetry operations which are not broken by the presence of the laser field. The real time propagation can be seen as an iterative solver of the BSE. It is less demanding than the exact diagionalization, and in a way more similar to perturbative schemes. The  drawback is that the number of iterations (i.e. time steps) may become very high. This depends on the goal of the propagation. To obtain the results described in the present manuscript only linear absorption properties are required and the cost is not much more demanding compared to standard BSE iterative solvers like Haydock. For the TR-ARPES spectral function however it is required to construct $G^<(t,t’)$ within GKBA. The implementation I did is based on storing to disk the density matrix history, which requires a lot of disk space. Alternative implementations which avoid this are however possible.

\subsection{Perturbative analysis of the EOM for the one-body density matrix}

In view of the analysis of the state generated by the laser pulse we expand the EOM for the density matrix $\rho$ up to second order in the pump laser pulse $\blE(t)$.
\begin{equation}
\rho_{nm\blk}(t) = \delta_{nm}f^{eq}_{n\blk} + \rho_{nm\blk}^{(1)}(t)+\rho_{nm\blk}^{(2)}(t) +\cdots \nonumber \, .
\end{equation}
For times preceding the activation of the pump ${\rho_{nm\blk}=\delta_{nm}f^{eq}_{n\blk}}$, i.e. $\rho$ is defined by the equilibrium occupations of states. 
Later, when discussing the transient absorption signal, we will consider the first order response in the probe starting from the state generated by the pump. We split the analysis in the independent particles (IP) part and in Hartree plus Screened EXchange part (HSEX). The goal of this expansion is to highlight the role of the intraband dipoles $\bld_{nn\blk}$, that are difficult to define and which we neglect in our numerical implementation.

\subsubsection{IP approximation}

I express the EOM at the IP level in the following form
\begin{equation}\label{eq:dm_eom}
i\partial_t\rho_{nm\blk} -(\Delta\epsilon_{nm\blk}-i\eta) \rho_{nm\blk}  = F_{nm\blk}[\rho] \, ,
\end{equation}
The source term of Eq.~\eqref{eq:dm_eom} reads
\begin{equation}\label{eq:dm_source}
F_{nm\blk}[\rho] = \blE(t)\cdot
\sum_{p}\left(\bld_{np\blk}\rho_{pm\blk}-\rho_{np\blk}\bld_{pm\blk}\right)
\, ,
\end{equation}
where the $\bld_{nm\blk}$ are the matrix elements of the dipole operator in the KS basis.
Expanding order by order equations
\eqref{eq:dm_eom} and \eqref{eq:dm_source} provides a chain of differential equations
for the $\rho^{(i)}_{nm\blk}$:
\begin{equation}\label{eq:dm_perturbative}
i\partial_t\rho^{(i)}_{nm\blk} -(\Delta\epsilon_{nm\blk}-i\eta)  \rho^{(i)}_{nm\blk}  =
F^{(i)}_{nm\blk} \, ,
\end{equation}
which can be hierarchically solved starting from the lowest order. At first order
\begin{equation}\notag
F^{(1)}_{nm\blk} = (f^{eq}_{n\blk}-f^{eq}_{m\blk})\blE(t)\cdot \bld_{nm\blk} \, ,
\end{equation}
is non vanishing only for transitions between valence and conduction states.
The solution of equation \eqref{eq:dm_perturbative} at first order shows that only
the matrix elements $\rho^{(1)}_{cv\blk}$ with 
$\Delta\epsilon_{cv\blk} \sim \omega_{pump}$ are activated by the pump through a
resonance mechanism. After an initial transient regime, these terms exhibit
an oscillatory behavior eventually suppressed by an exponential factor, \emph{i.e.}
$\exp(-i(\Delta\epsilon_{cv\blk}-i\eta_{cv\blk})t)$ if decoherence is accounted for.
Going ahead in the hierarchy of equations \eqref{eq:dm_perturbative}, the ${\rho^{(2)}_{nm\blk}}$ terms are activated. Using the fact that $\rho^{(1)}_{cc'\blk}=\rho^{(1)}_{vv'\blk}=0$ we can simplify the second order term.
In the $cc'$ and $vv'$ channel the source term associated to transitions among conduction states reads
\begin{eqnarray}
F^{(2)}_{cc'\blk} =
\blE(t)\cdot\sum_{v}\left[\mathbf{d}_{cv\blk}\rho_{vc'\blk}^{(1)}-\rho_{cv\blk}^{(1)}\mathbf{d}_{vc'\blk}\right] \, ,  \label{eq:dm_2nd_order_cc} \\ 
F^{(2)}_{vv'\blk} =
\blE(t)\cdot\sum_{c}\left[\mathbf{d}_{vc\blk}\rho_{cv'\blk}^{(1)}-\rho_{vc\blk}^{(1)}\mathbf{d}_{cv'\blk}\right] \, .  \label{eq:dm_2nd_order_vv}  
\end{eqnarray}
The $\rho^{(2)}_{cc'\blk}$ ($\rho^{(2)}_{vv'\blk}$) terms are activated when both $\Delta\epsilon_{cv\blk}$ and $\Delta\epsilon_{c'v\blk}$ ($\Delta\epsilon_{cv'\blk}$) lay within $\omega_0-\Delta_0$ and
$\omega_0+\Delta_0$. 
In the $cv$ channel
\begin{equation}
F^{(2)}_{cv\blk} =
\blE(t)\cdot\left[ \sum_{c'}\mathbf{d}_{cc'\blk}\rho_{c'v\blk}^{(1)}-\sum_{v'}\rho_{cv'\blk}^{(1)}\mathbf{d}_{v'v\blk}\right] \, ,  \label{eq:dm_2nd_order_cv} 
\end{equation}
which shows the importance of intra-band dipoles appearing already at second order in the coupling with the external field. In the implementation of the \texttt{yambo\_rt} code, dipoles are set to zero whenever $\Delta\epsilon_{nm\blk}<\epsilon_{thresh}$ (and in particular  for $n=m$), due to the fact that the dipoles $\mathbf{d}_{nm\blk}$ are ill defined in such cases. More in general, neglecting all valence--valence and conduction--conduction matrix elements we obtain the No--Intraband (NI) term
\begin{equation}
F^{(2NI)}_{cv\blk}=0
\end{equation}
and, due to boundary conditions, ${\rho^{(2)}_{cv\blk}=0}$.
Since we are interested in the leading order terms in the pump pulse $\rho_{cv\blk}^{(1)}$ and $\rho^{(2)}_{cc'\blk}$ we can ignore the role of intra-band dipoles for weak pumps. 

\subsubsection{HSEX approximation}

We now want to check how much the update of the HSEX self-energy influences the above analysis. $\Delta\Sigma^{HSEX}$ is linear in $\rho$ (since we do not update the screening)
\begin{equation}
\Delta\Sigma^{HSEX}[\rho(t)]_{nm\blk}= K^{HSEX}_{nm\blk,\nt\mt\blkt}\rho_{\nt\mt\blkt},
\end{equation}
where $K^{HSEX}=\partial\Sigma^{HSEX}/\partial G$ is the standard \ai-BSE kernel, however the indexes are, in general, not restricted to the $cv$ channel.
$\Delta\Sigma^{HSEX}$ enters the EOM in a term of the form $Q^{HSEX}=[\Delta\Sigma^{HSEX},\rho]$.
The leading order term is different from zero only in the cv channel
\begin{equation}
Q^{(1)}_{cv\blk}= (f^{eq}_{v\blk}-f^{eq}_{c\blk}) K^{HSEX}_{cv\blk,\ct\vt\blkt}\rho^{(1)}_{\ct\vt\blkt},
\end{equation}
This terms gives a correction into the EOM
\begin{equation}
\Delta\epsilon_{cv\blk}\rho^{(1)}_{cv\blk}\rightarrow
\left(\Delta\epsilon_{cv\blk}\delta_{v,\vt}\delta_{c,\ct}\delta(\blk-\blkt)+(f^{eq}_{v\blk}-f^{eq}_{c\blk})K^{HSEX}_{cv\blk,\ct\vt\blkt}\right)\rho^{(1)}_{\ct\vt\blkt}
\end{equation}
and gives a renormalization of the energies from $\Delta\epsilon_{cv\blk}$ to the poles of the equilibrium BSE at zero momentum $\omega_{\lambda\mathbf{0}}$.
As a consequence the state generated by the pump oscillates at the excitonic energies.

%
The second order term has the form
\begin{equation}
Q^{(2)}_{nm\blk}= K^{HSEX}_{n\lt\blk,\nt\mt\blkt}\rho^{(1)}_{\nt\mt\blkt}\rho^{(1)}_{\lt m\blk}-\rho^{(1)}_{n\lt\blk}K^{HSEX}_{\lt m\blk,\nt\mt\blkt}\rho^{(1)}_{\ct\vt\blkt}+
(f^{eq}_{m\blk}-f^{eq}_{n\blk}) K^{HSEX}_{nm\blk,\nt\mt\blkt}\rho^{(2)}_{\nt\mt\blkt}
\label{eq:G2_general}
\end{equation}
This couples the $cv$ channel with the $cc'$ and $vv'$ channels.
We can write this term explicitly in the $cc'$ and $vv'$ channels:
\begin{eqnarray}
Q^{(2)}_{cc'\blk}&=& \left(K^{HSEX}_{c\vt'\blk,\ct\vt\blkt}\rho^{(1)}_{\vt'c'\blk}-\rho^{(1)}_{c\vt'\blk}K^{HSEX}_{\vt'c'\blk,\ct\vt\blkt}\right)\rho^{(1)}_{\ct\vt\blkt} \\
Q^{(2)}_{vv'\blk}&=& \left(K^{HSEX}_{v\ct'\blk,\ct\vt\blkt}\rho^{(1)}_{\ct'v'\blk}-\rho^{(1)}_{v\ct'\blk}K^{HSEX}_{\ct'v'\blk,\ct\vt\blkt}\right)\rho^{(1)}_{\ct\vt\blkt} 
\end{eqnarray}
where the last term of eq.~\eqref{eq:G2_general} gives no contribution. Accordingly
$\rho^{(2)}_{cc'\blk}$ and $\rho^{(2)}_{vv'\blk}$ become independent from the terms $\rho^{(2)}_{cv\blk}$. This means that, neglecting the intra-band dipoles, $\rho^{(2)}_{cc'\blk}$ and $\rho^{(2)}_{vv'\blk}$ remains exact up to second order also at the TD-HSEX level. Moreover looking at these terms for the case $c=c'$ we understand why, at the HSEX level, $N_c=\sum_c\rho^{(2)}_{cc\blk}$ displays fast oscillations as long as $\rho_{cv\blk}\neq 0$, as shown in the manuscript.

\subsection{Generated state in the continuum}  \label{ssec:pumping_continuum}

\begin{figure}[h]
	\includegraphics[width=0.45\textwidth]{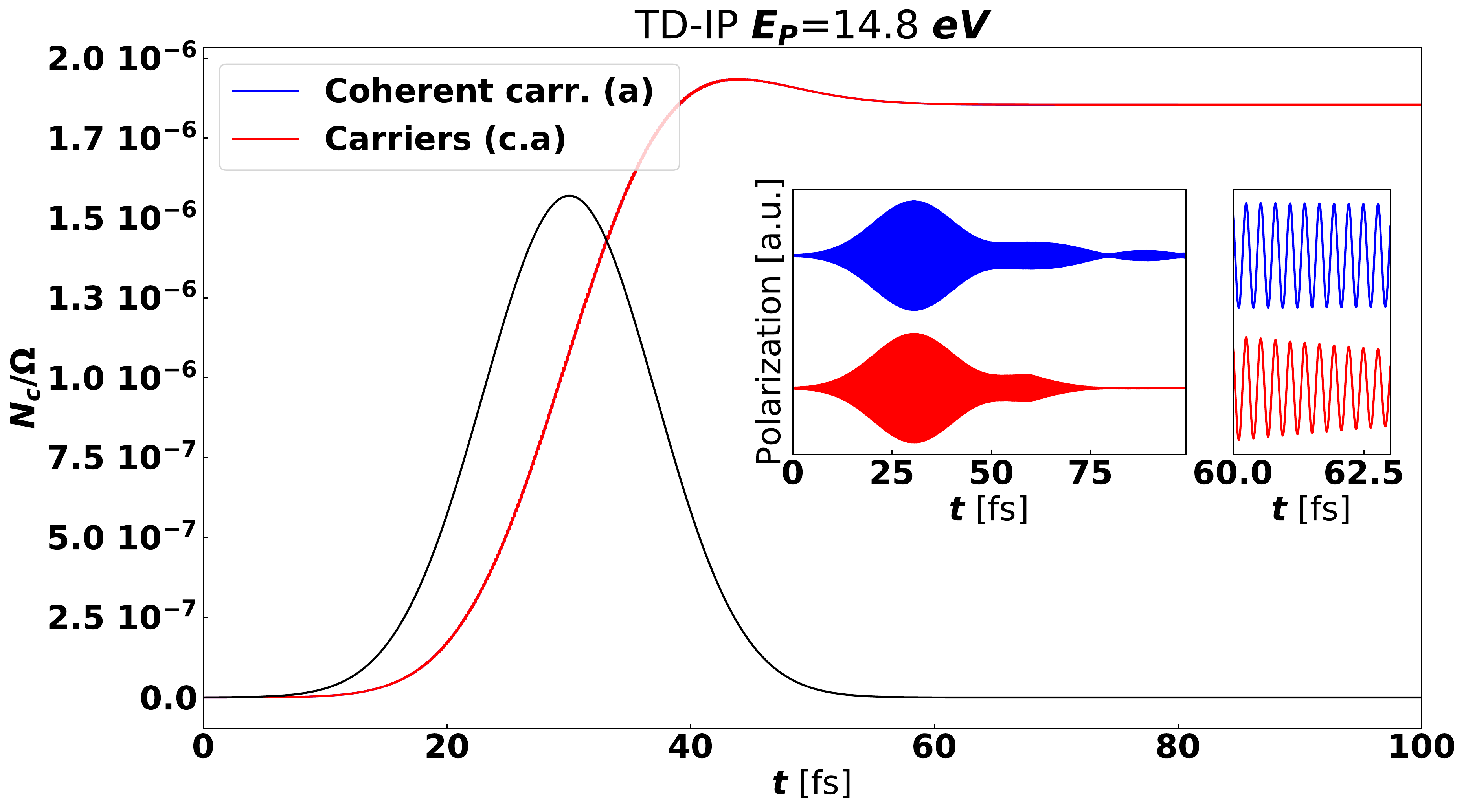}
	\includegraphics[width=0.45\textwidth]{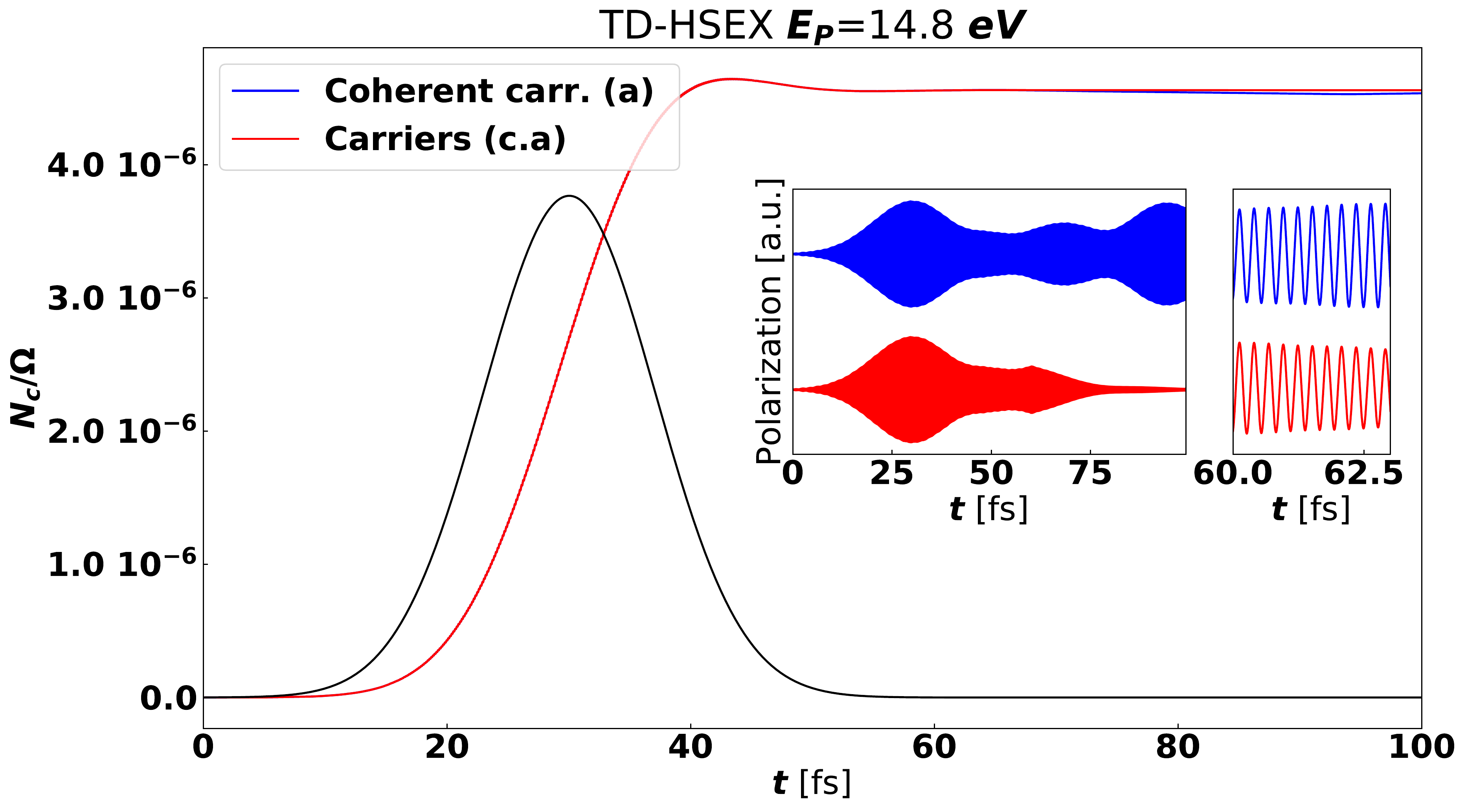}
	\caption{Time dependent polarization obtained propagating the equation of motion for $\rho(t)$, with the laser pulse in the continuum at the IP level (left panel) and at the HSEX level (right panel).} 
	\label{fig:states_cont}
\end{figure}

We now analyze the state generated with the laser pulse frequency in the continuum, comparing the result of the TD-IP propagation with a full TD-HSEX propagation. As shown in Fig.~\ref{fig:states_cont} the total poplation display a similar behavior, although the total number of carriers generated is higher (by roughly a factor 2) in the TD-HSEX case. This is due to the fact that the TD-HSEX (i.e. BSE) absorption is higher than the IP absorption in the same energy range. We will analyze more in details the distribution of the population with the TR-ARPES signal. Here we also observe that, at the IP level, the residual polarization is lower since more frequencies exist at the IP level in the energy range considered by the pump pulse.

\clearpage

\section{Time Resolved ARPES singal}  

\subsection{Construction of the $G^<(t,t')$ Green function}  \label{ssec:G_lesser_GKBA}

The GKBA expression for $G^<$ is
\begin{align}
G^<_{cc'\blk}(t,t')=\rho_{c\nt\blk}(t)G^{(a)}_{\nt c'\blk}(t,t')-G^{(r)}_{c\nt\blk}(t,t')\rho_{\nt c'\blk}(t').
\label{eq:G_lesser_cc_GKBA}
\end{align}
The retarded and the advanced propagator at the TD-HSEX level are 
given by
\begin{eqnarray}
G^{(r)}(t,t')&=& -i\theta(t-t') T\[e^{-i\int_{t'}^t 
	h^{HSEX}[\rho(t)]dt}\].
\end{eqnarray}
The propagator in practice is constructed as
\begin{eqnarray}
G^{(r)}(t,t')&=& -i\theta(t-t') \prod_{n=0}^N e^{-ih^{HSEX}[\rho(t-n\,dt)]dt} 
\end{eqnarray}
with $t-N\,dt=t'$. Starting from the time step $t=t'$ where $G^{(r)}_{nm\blk}(t,t)=-i\delta_{nm}$ we have
\begin{eqnarray}
G^{(r)}(t+dt,t'-dt) = e^{-i h^{HSEX}[\rho(t+dt)]dt} G^{(r)}(t,t') e^{-i h^{HSEX}[\rho(t'-dt)]dt}
\end{eqnarray}
In the actual implementation I replace
\begin{eqnarray}
h^{HSEX}[\rho(t+dt)] &\rightarrow& 0.5\,(h^{HSEX}[\rho(t+dt)]-h^{HSEX}[\rho(t)]) \\
h^{HSEX}[\rho(t'-dt)] &\rightarrow& 0.5\,(h^{HSEX}[\rho(t'-dt)]-h^{HSEX}[\rho(t')]).
\end{eqnarray}
The exponential $e^{-ih^{HSEX}[\rho(t)]dt}$ is approximated with a 5th order expansion in $dt$ and, for each term on the left and on the right, a matrix-matrix multiplication in the $nm$ space is performed. This is repeated for each $\blk$-point for each time-step.
The Hamiltonian is defined as (we consider only the non overlapping regime, i.e. we probe time ranges where the pump pulse is off)
\begin{equation}
h^{HSEX}[\rho(t)]=h^{eq}+\Delta\Sigma^{HSEX}[\rho(t)].
\end{equation}
In general
\begin{equation}
\Delta\Sigma^{HSEX}[\rho(t)]_{nm\blk}=K^{HSEX}_{nm\blk,\nt\mt\blkt}(\rho_{\nt\mt\blkt}(t)-\rho^{eq}_{\nt\mt\blkt}).
\end{equation}
The sum over ${\nt\mt\blkt}$ indexes is implicit.

To analyze the expression, we write again the perturbative expansion (this is not done in the implementation).
As discussed previously $\rho_{cc'\blk}(t)\approx\rho^{(2)}_{cc'\blk}(t)$ (in the $vv'$ channel instead $\rho_{vv'\blk}(t)\approx\delta_{v,v'}f^{(eq)}_{v\blk}$), while
$\rho_{cv\blk}(t)\approx\rho_{cv\blk}^{(1)}(t)$.
For the self-energy and we consider the leading order terms only. If we restrict to the $cv$ channel,
\begin{equation}
\Delta\Sigma^{HSEX}[\rho(t)]_{cv\blk}\approx K^{HSEX}_{cv\blk,\ct\vt\blkt}\rho^{(1)}_{\ct\vt\blkt}.
\end{equation}
where $K^{HSEX}$ is the standard \ai-BSE kernel. The diagonal terms $\Sigma_{nn\blk}$ can be absorbed in a renormalization of the energies entering the $h^{eq}$ hamiltonian.
The propagators have leading terms $G^{(r/a)}_{nn\blk}(t,t')\approx G^{(r/a),(eq)}_{nn\blk}(t,t')$ and $G^{(r/a)}_{cv\blk}(t,t')\approx G^{(r/a),(1)}_{cv\blk}(t,t')$.
The full $G^<$ can be split into two terms, one with contribution from the terms of the density matrix $\rho_{cc'\blk}(t)$ and one with contribution from the off-diagonal elements $\rho_{cv\blk}(t)$
Interestingly both terms are of second order in the pump pulse:
\begin{eqnarray}
G^{<,(2a)}_{cc'\blk}(t,t')&=&\rho^{(2)}_{c\ct\blk}(t)G^{(a),(eq)}_{\ct c'\blk}(t,t')-G^{(r),(eq)}_{c\ct\blk}(t,t')\rho^{(2)}_{\ct c'\blk}(t'), \label{eq:G_lesser_cc_GKBA_cc} \\
G^{<,(2b)}_{cc'\blk}(t,t')&=&\rho^{(1)}_{c\vt\blk}(t)G^{(a),(1)}_{\vt c'\blk}(t,t')-G^{(r),(1)}_{c\vt\blk}(t,t')\rho^{(1)}_{\vt c'\blk}(t'), 
\label{eq:G_lesser_cc_GKBA_cv}
\end{eqnarray}
The signal for the case $\rho_{cv\blk}=0$ is fully determined by eq.~\eqref{eq:G_lesser_cc_GKBA_cc}, while the complete signal has also contribution from eq.~\eqref{eq:G_lesser_cc_GKBA_cv}. It is remarkable to notice that, in the coherent exciton case, the contribution eq.~\eqref{eq:G_lesser_cc_GKBA_cv} not only gives the excitonic ARPES singal in the band gap, but is also responsible for exactly removing the qp ARPES singal from the conduction band due to eq.~\eqref{eq:G_lesser_cc_GKBA_cc}.

\begin{figure}[h]
	\includegraphics[width=0.45\textwidth]{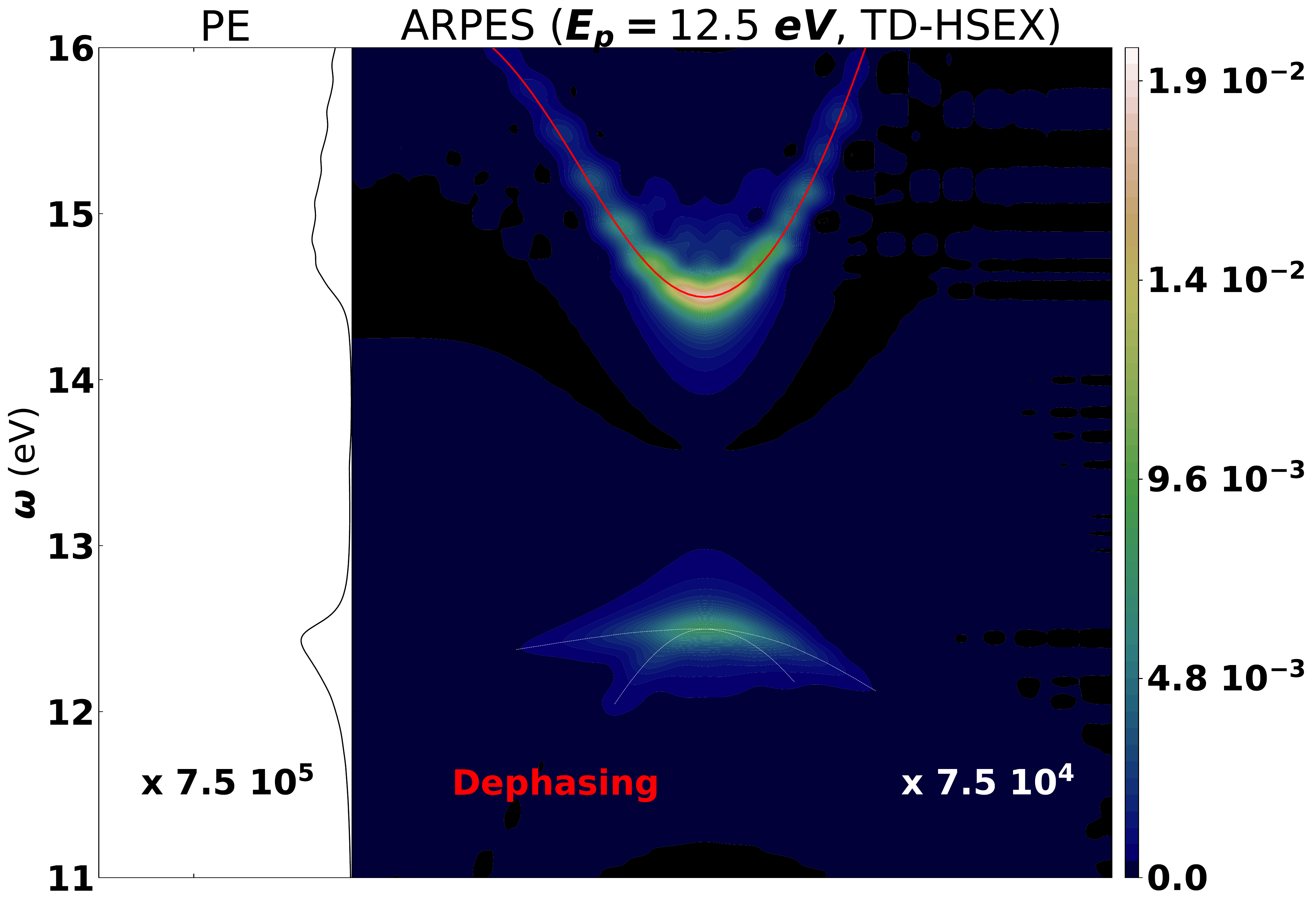}
	\caption{The TR-ARPES spectra function after partial de-phasing of the off-diagonal elements of the density matrix. The numbers reported in the PE and ARPES panels represent the magnification factor needed for the signal in the conduction band region to reach similar intensity as the signal from the valence band.} 
	\label{fig:TD-HSEX_exc_deph_partial}
\end{figure}

Here we also show (see Fig.~\ref{fig:TD-HSEX_exc_deph_partial}) the situation in which the matrix elements of the density matrix are partially dephased. In such case the cancellation between the two contributions to $G^<_{cc\blk}$ is not exact anymore and both the signal from the excitonic state and from the carriers are visible. This is the intermediate situation between the coherent excitonic case and the non coherent carriers population shown in the manuscript. It is reported here to highlight the importance of the cancellation between such two terms. As discussed in the main text the procedure of sending to zero the off-diagonal elements of the density matrix changes the total energy of the system and it is thus not physical. This is somehow evident also from the ARPES spectral function since lower photon energy is required to extract electrons after these elements are sent to zero.

%

\subsubsection{Interpolation of $G^<(t,t')$ in $\blk$-space}

\begin{figure}[h]
	\includegraphics[width=0.45\textwidth]{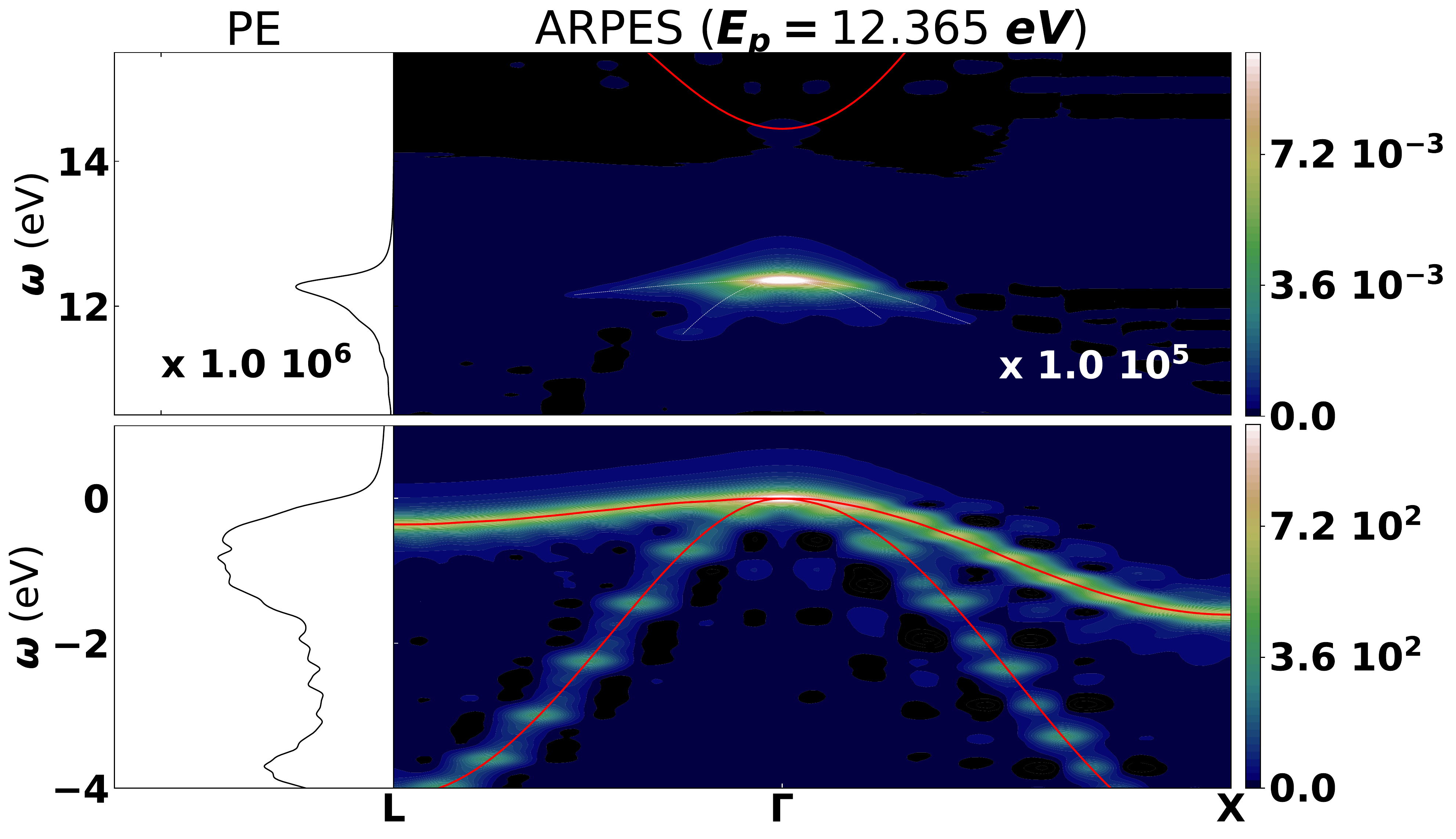}
	\includegraphics[width=0.45\textwidth]{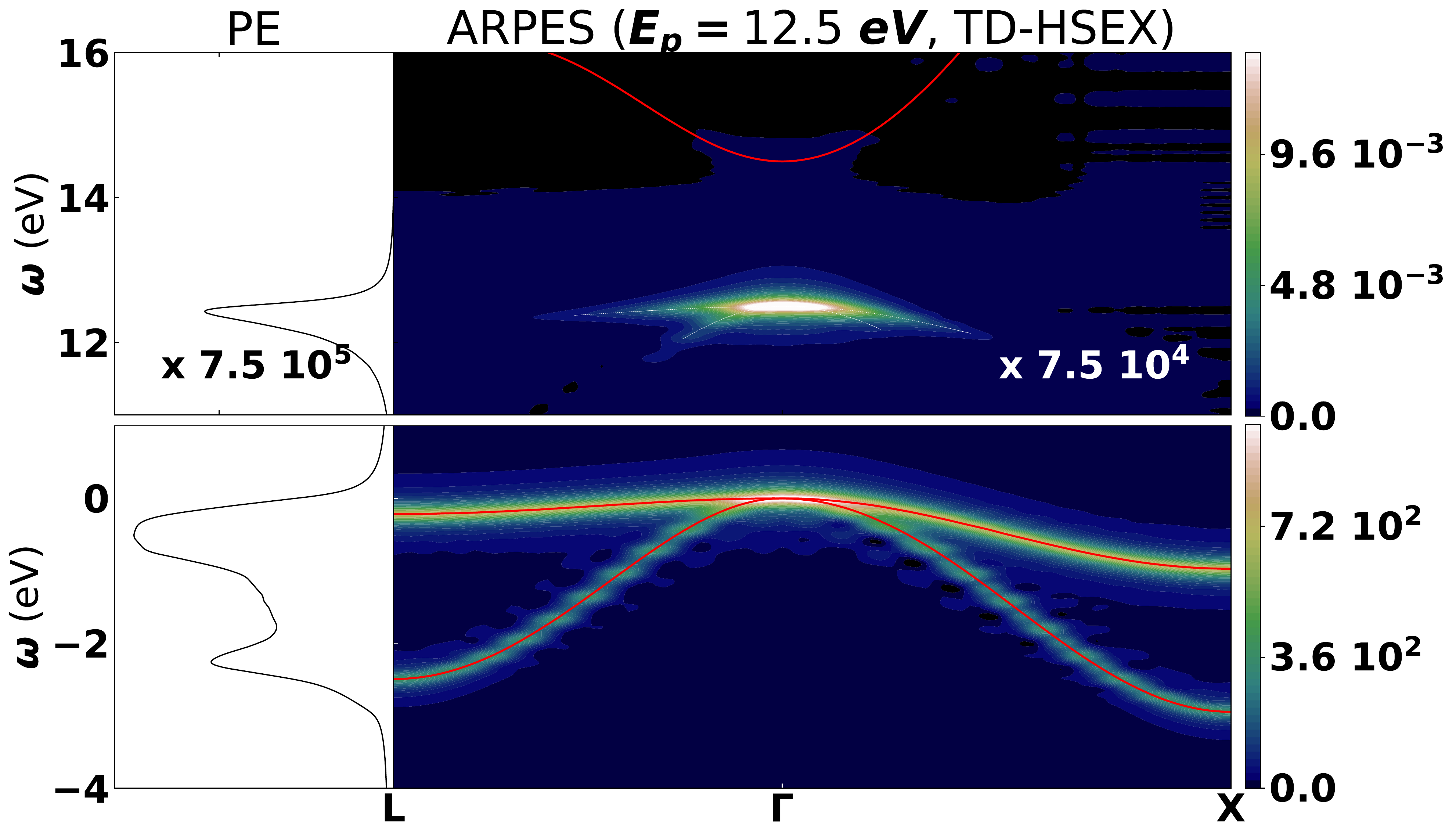}
	\caption{Integrated and $k$-resoved spectral function for LiF after the action of an ultra short laser pulse with frequency resonant to the excitonic peaks.
		Two different k-point samplings are used $G^<(\omega)$. The numbers reported in the PE and ARPES panels represent the magnification factor needed for the signal in the conduction band region to reach similar intensity as the signal from the valence band.} 
	\label{fig:TR-ARPES_interpolation}
\end{figure}

The $G^<_{cc\blk}(\omega)$ function is constructed after the TD-HSEX propagation on a regular grid in $\blk$-space. To plot the signal on a given path a Fourier interpolation is performed (this is possible since the physical signal is phase independent). However the interpolation of a function is highly non trivial. We solve this doing a frequency by frequency interpolation. This is much less efficient since the resulting signal tends to be stretched along the fixed frequency axis. In this case it works reasonably well since a very fine $\blk$-points sampling is needed to converge the absorption spectrum.

In Fig.~\ref{fig:TR-ARPES_interpolation} we show the interpolated signal starting from the $16\times 16\times 16$ sampling (left panel) and from the $24\times 24\times 24$. The excitonic signal is very smooth already with the smaller grid, while the signal from the valence band from the $16^3$ grid suffers from the stretching above mentioned.

\subsection{TR-ARPES signal in the continuum}  \label{ssec:ARPES_continuum}

\begin{figure}[h]
	\includegraphics[width=0.33\textwidth]{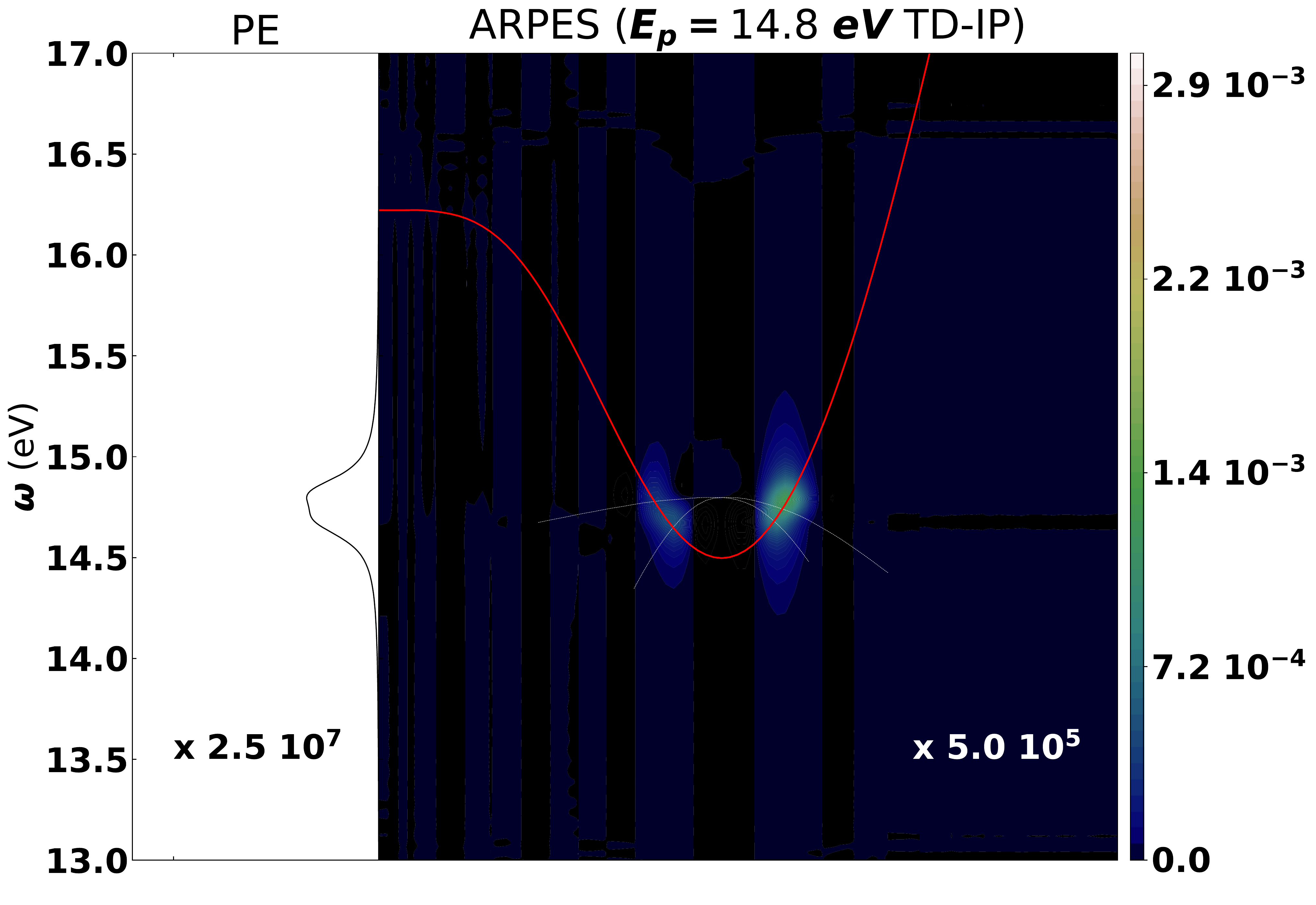}
	\includegraphics[width=0.33\textwidth]{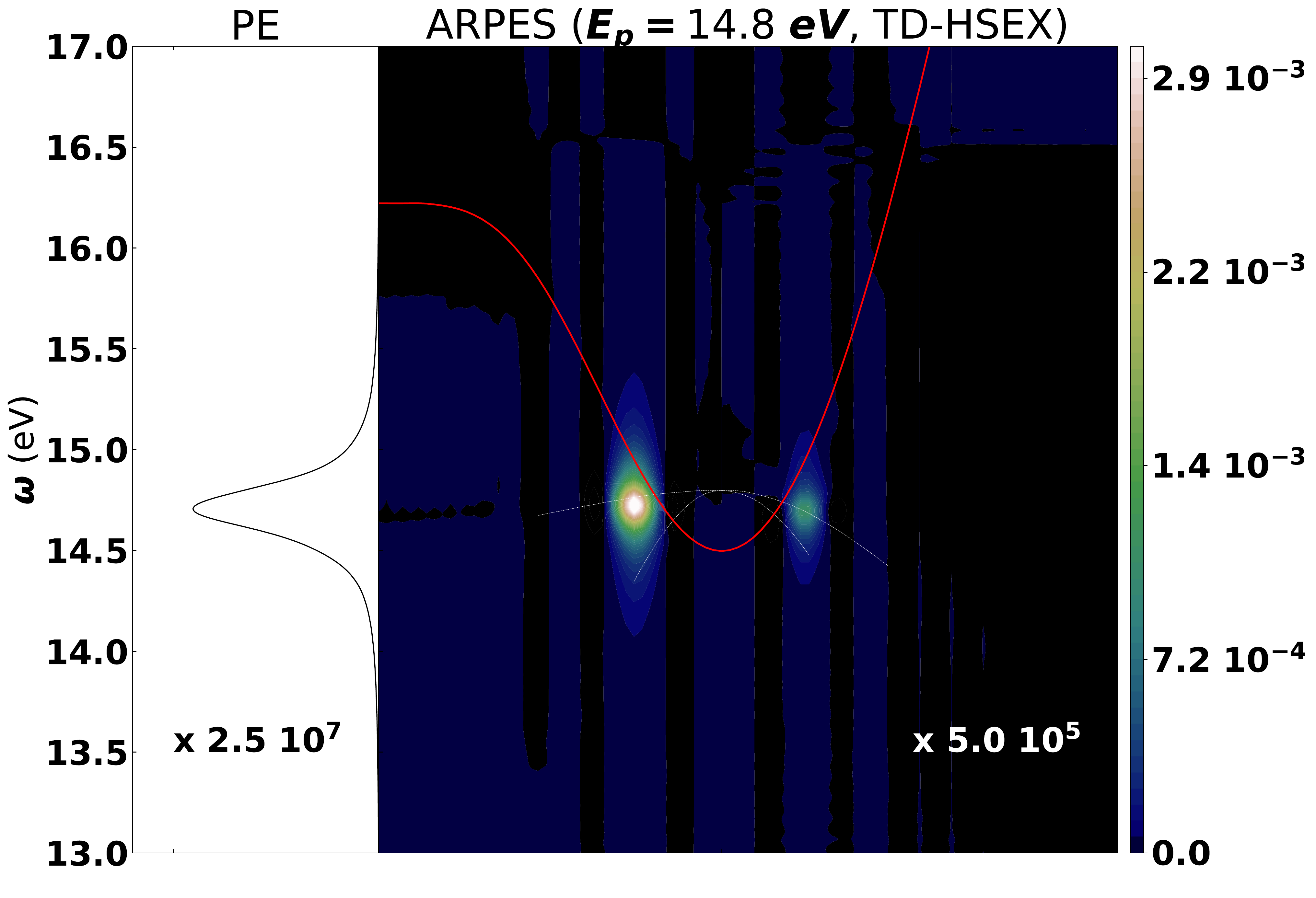}
	\includegraphics[width=0.33\textwidth]{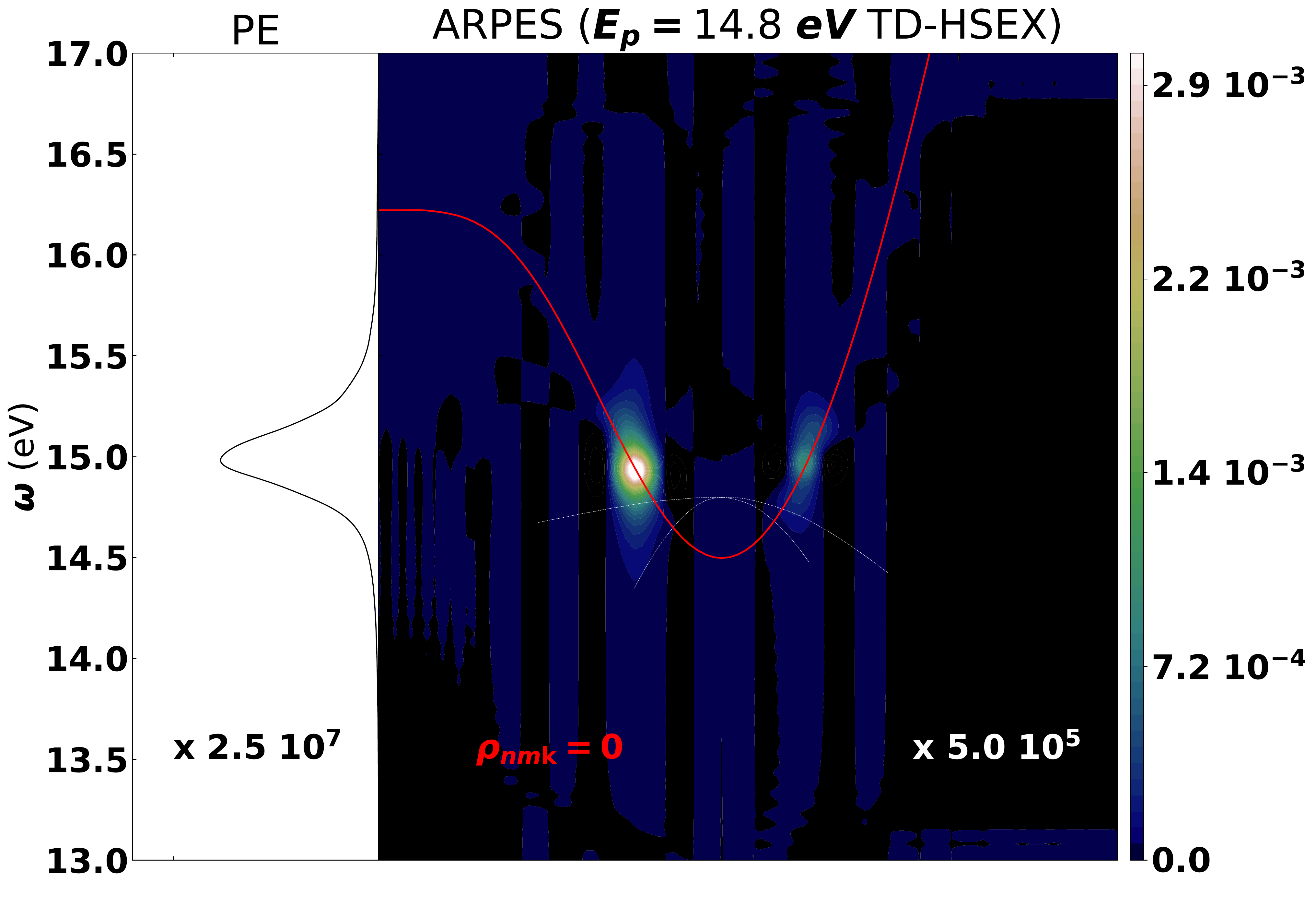}
	\caption{Integrated and $k$-resoved spectral function for LiF after the action
		of an ultra short laser pulse in the continuum.
		The $k$-resoved signal is obtained interpolating $G^<(\blk,\omega)$ frequency by frequency
		from a regular grid. The band structure is superimposed to the signal (red lines).
		A replica of the valence band shifted by the excitonic energy is also represented
		with white lines. The three panels represent TD-IP simultion (left panel) TD-HSEX simulation (central panel), and TD-HSEX simulation with dephasing (right panel).} 
	\label{fig:PE_and_ARPES_cont}
\end{figure}

In Fig.~\ref{fig:PE_and_ARPES_cont} we compare the TR-ARPES signal from a system for the case $\omega_{P_a}=14.8$~eV at the TD-IP and the TD-HSEX level. For the TD-HSEX level we consider both the the result of the simulation without dephasing and the result after sending to zero the off-diagonal matrix-elements of the density matrix.
The obtained signal is almost identical in the three cases, with two peaks located at the point where the up-shifted valence band crosses the conduction band.
As for the case with the laser pulse resonant with the exciton, the TD-HSEX solution predicts that the signal
is a replica of the valence band weighted by the BSE eigenvector which has in principle no direct relation to the
conduction band. However, in this energy range, the BSE eigenvectors in the continuum $A^{\lambda\blq}$ are of almost delta like nature for LiF, i.e. $A^{\lambda\blq}\approx \delta_{v,v_{P_a}} \delta_{c,c_{P_a}} \delta(\blk-\blk_{P_a})$ and the expression of the ARPES spectral function reduces to the IP expression with $\epsilon_{c\blk}=(\epsilon_{v\blk-\blq}+\omega_{\lambda\blq})$. Numerically this results can be approached only very slowly as discussed above.
The results from this section show that our approach is completely general and it works well in  the excitonic case, but also in the continuum where however a simpler independent particles approach and interpretation in
terms of carriers moving on the band structure works fine.

There is a small shift in $\blk$--space between the IP and HSEX approach (panel $a$ and panel $b$)
which turns into an energy shift once dephasing is included in the TD-HSEX simulation.
The source of the shifts in Fig.~\ref{fig:PE_and_ARPES_cont} can be understood considering these points:
\begin{itemize}
	\item[(i)]{ the energy of the photo--emitted signal must be the energy from the valence band,
		plus the energy of the laser pump $\epsilon_{v\blk}+\omega_0$
		if we do not artificially change the energy of the system (panels $a$ and $b$)}
	\item[(ii)]{ the energy of the photo--emitted electron must lay on the conduction band
		structure when it depends only on $G^{<,2a}_{cc\blk}(\omega)$
		(panels $a$ and $c$)}
	\item[(iii)]{ the dephasing alters the energy of the system.
		The energy of the photo--emitted electrons is shifted from $\epsilon_{v\blk}+\omega_0$ with $\omega_0 \approx E^{\lambda}$,
		to $\epsilon_{c\blk}$, i.e. the same shift discussed in the analysis of $F^{\lambda}$.
		However it does not change where the signal lies in $\blk$--space, since it just kills
		$G^{<,2b}_{cc\blk}(\omega)$, thus the signal from TD-HSEX simulation always lies in the same
		$\blk$--region (panels $a$ and $b$).}
\end{itemize}
We expect all these shifts to go to zero at convergence, and thus regard them as non-physical. However this should be fully verified. The existence of this shifts is the main reason why in the main text we prefer to present the signal obtained from within the IP propagation for the case  in the continuum.

\begin{figure}[b]
	\includegraphics[width=0.33\textwidth]{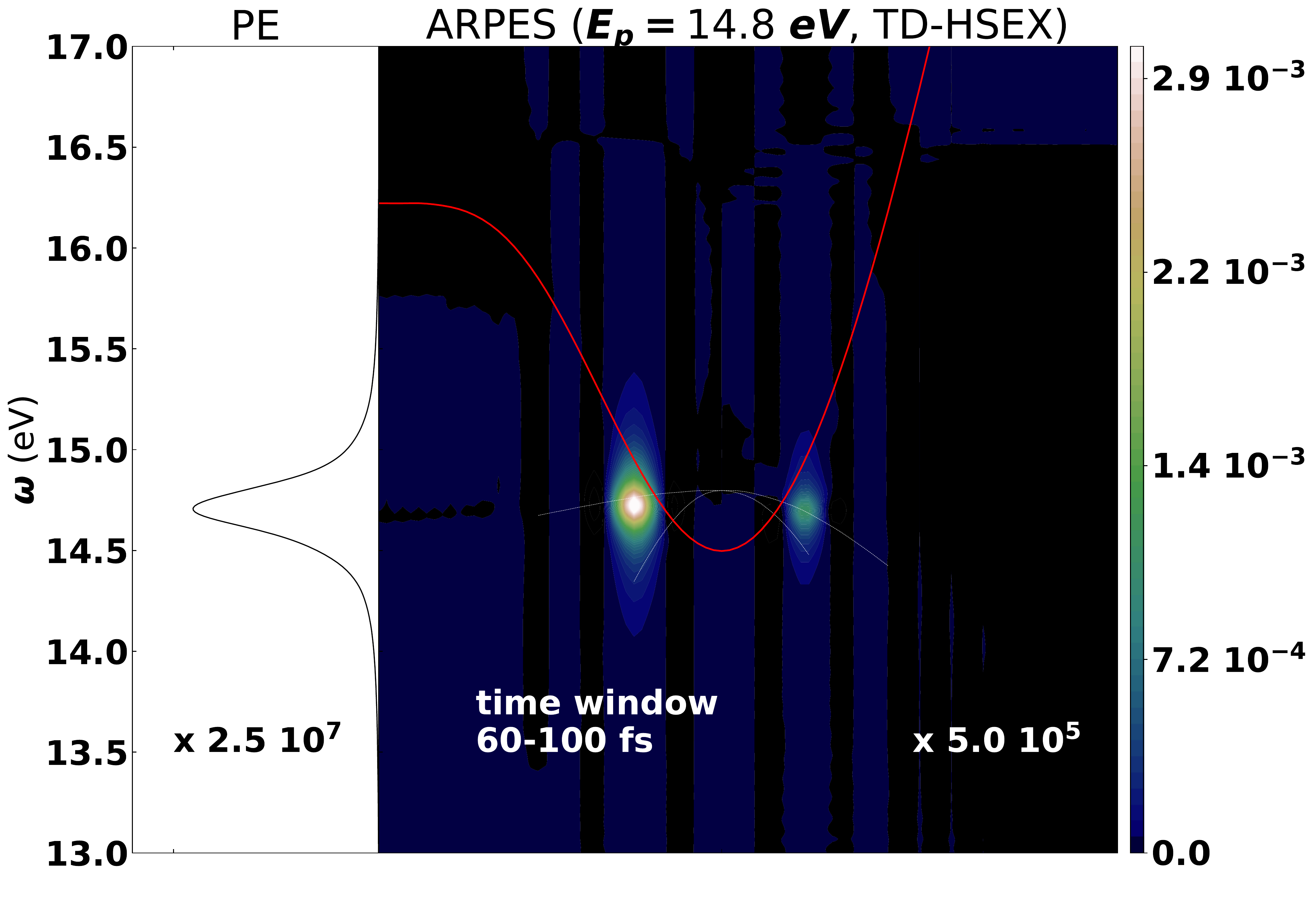}
	\includegraphics[width=0.33\textwidth]{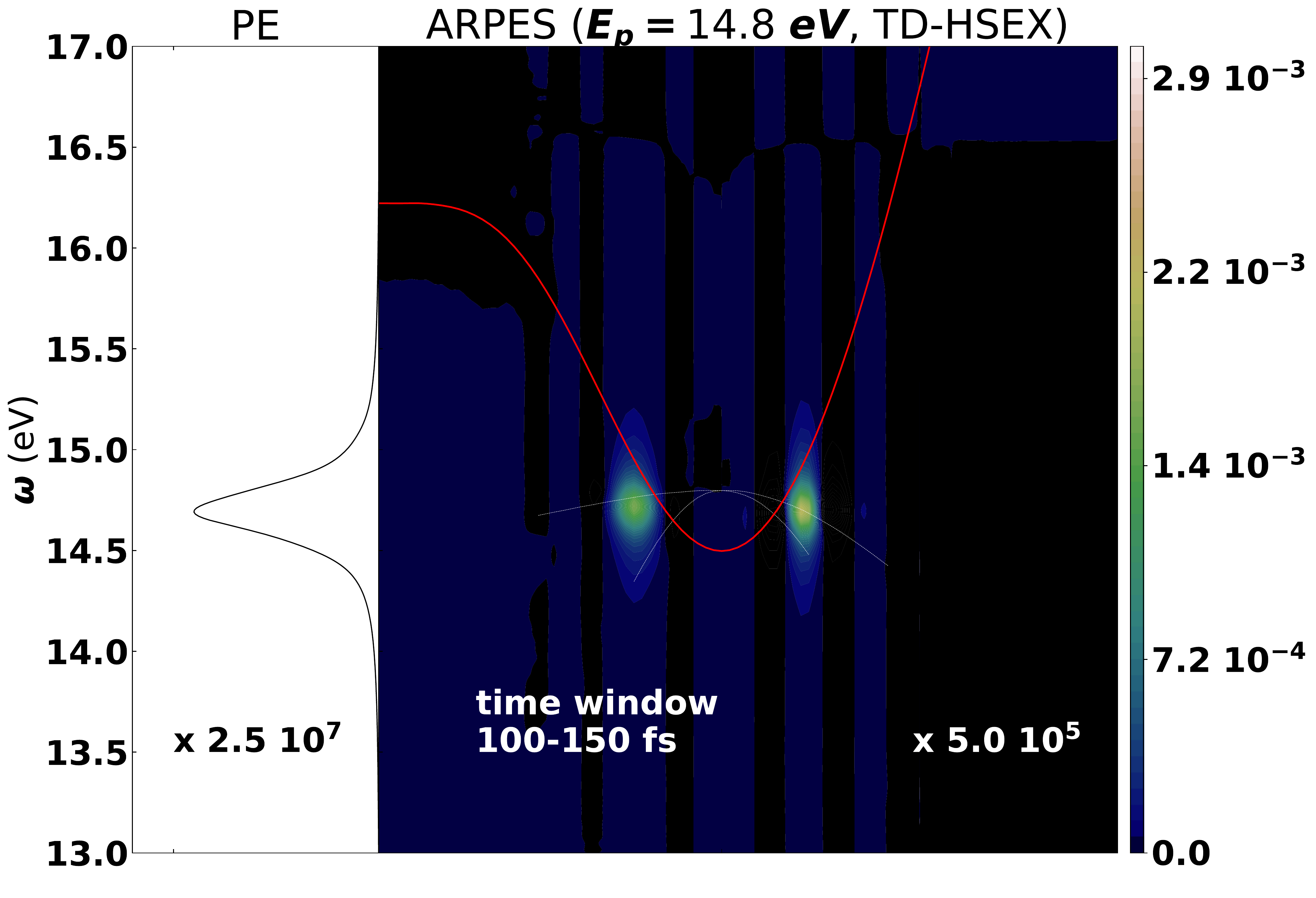}
	\caption{Integrated and $k$-resoved spectral function for LiF after the action of an ultra short laser pulse with frequency in the continuum (bottom panels).
		Two different time windows are considered to reconstruct $G^<(\omega)$.} 
	\label{fig:TD-HSEX_pol_cont}
\end{figure}

We conclude our analysis looking at the relative intensity of the two peaks.
Despite we are in the continuum with a very high density of poles in our simulation, only few are optically active
(see Fig.~\ref{fig:ABS_convergence}). This is reflected in the behavior of the polarization, as shown in Fig.~\ref{fig:TD-HSEX_pol_cont}.
This points to the fact that, at variance with the case where we pump resonant with the excitonic frequency,
the PE signal may depend on the time window considered to construct it.
To address this we consider the spectrum generated into two different energy windows: $160-200 fs$ and
$200-250 fs$. Both spectra consist of two peaks located in the same point. 
However, by changing the time window used to extract the spectral function, the relative intensity of the two peaks changes (see fig.~\ref{fig:TD-HSEX_pol_cont}).

\clearpage

\section{Transient Absorption signal}  \label{sec:TrAbs}

\subsection{Perturbative expansion in the probe}  \label{ssec:probe_pert}

\subsubsection{IP approximation} \label{sssec:IP_approx}

We now proceed with the perturbarbative expansion defined in the previous section.
We are interested in the linear response to the probe $\ble(t)$. Since the signal generated by the pump pulse has mixed first and second order character, the TrAbs signal will have in general mixed second and third order character. After the action of the pump pulse the density matrix contains both diagonal and off-diagonal elements and the third order source terms have the general expression.
In the $cc'$ and $vv'$ channels we have
\begin{eqnarray}
F^{(1p)}_{cc'\blk} &=&
\ble(t)\cdot\Bigg[\sum_{v}\left(\mathbf{d}_{cv\blk}\rho_{vc'\blk}^{(1+2)}-\rho_{cv\blk}^{(1+2)}\mathbf{d}_{vc'\blk}\right)+
\sum_{c''}\left(\mathbf{d}_{cc''\blk}\rho_{c''c'\blk}^{(2)}-\rho_{cc''\blk}^{(2)}\mathbf{d}_{c''c'\blk}\right) \Bigg] \, .  \label{eq:dm_1st_order_probe_cc}   \\
F^{(1p)}_{vv'\blk} &=&
\ble(t)\cdot\Bigg[\sum_{c}\left(\mathbf{d}_{vc\blk}\rho_{cv'\blk}^{(1+2)}-\rho_{vc\blk}^{(1+2)}\mathbf{d}_{cv'\blk}\right)+
\sum_{v''}\left(\mathbf{d}_{vv''\blk}\rho_{v''v'\blk}^{(0+2)}-\rho_{vv''\blk}^{(0+2)}\mathbf{d}_{v''v'\blk}\right) \Bigg] \, .  \label{eq:dm_1st_order_probe_vv}  
\end{eqnarray}
In the $cv$ channel
\begin{equation}
F^{(1p)}_{cv\blk} =
\ble(t)\cdot\Bigg[
\sum_{c'}\left(\mathbf{d}_{cc'\blk}\rho_{c'v\blk}^{(1+2)}-\rho_{cc'\blk}^{(2)}\mathbf{d}_{c'v\blk}\right)+
\sum_{v'}\left(\mathbf{d}_{cv'\blk}\rho_{v'v\blk}^{(0+2)}-\rho_{cv'\blk}^{(1+2)}\mathbf{d}_{v'v\blk}\right)\Bigg] \, ,  \label{eq:dm_1st_order_probe_cv} 
\end{equation}
In both channels intra-band terms are needed. Not only the coupling with the external pulse needs to include them, but also the dipoles are needed in the definition of the polarization $\blP=\sum_{nm} \bld_{nm\blk} \rho_{nm\blk}$ from which the TrAbs signal is defined.

In the continuum the oscillatory terms quickly dephase and one can neglect them. This gives:
\begin{eqnarray}
F^{(1p-DE)}_{cc'\blk} &=&
\ble(t)\cdot\Bigg[
\sum_{c''}\left(\mathbf{d}_{cc''\blk}\rho_{c''c'\blk}^{(2)}-\rho_{cc''\blk}^{(2)}\mathbf{d}_{c''c'\blk}\right) \Bigg] \, .  \label{eq:dm_1st_order_probe_DE_cc}   \\
F^{(1p-DE)}_{vv'\blk} &=&
\ble(t)\cdot\Bigg[
\sum_{v''}\left(\mathbf{d}_{vv''\blk}\rho_{v''v'\blk}^{(0+2)}-\rho_{vv''\blk}^{(0+2)}\mathbf{d}_{v''v'\blk}\right) \Bigg] \, .  \label{eq:dm_1st_order_probe_DE_vv}  \\
F^{(1p-DE)}_{cv\blk} &=&
\ble(t)\cdot\Bigg[
\sum_{c'}\left(-\rho_{cc'\blk}^{(2)}\mathbf{d}_{c'v\blk}\right)+
\sum_{v'}\left(\mathbf{d}_{cv'\blk}\rho_{v'v\blk}^{(0+2)}\right)\Bigg] \, ,  \label{eq:dm_1st_order_probe_DE_cv} 
\end{eqnarray}
Thus the source term for the density matrix dynamics, which in turn determines the transient absorption signal, depends on non--equilibrium occupations only. The zero order term $\rho_{v'v\blk}^{(0)}$ is removed by subtracting the equilibrium contribution. The $cc'$ and $vv'$ terms still give contribution to the polarization. However neglecting the dipoles whenever $\Delta\epsilon_{nm\blk}<\epsilon_{thresh}$ just removes the low frequency terms, since $\rho_{cc'\blk}\propto e^{-i\Delta\epsilon{cc'\blk}}$ (and similarly $\rho_{vv'\blk}\propto e^{-i\Delta\epsilon{vv'\blk}}$) and the transient absorption for $\omega>\Delta\epsilon_{thresh}$ remains unaffected.

In the situation where, instead, there is an isolated pole, the $\rho_{cv\blk}$ terms can survive for longer times and the complete expression for the source terms need to be used. Neglecting the intra-band dipoles (and using, accordingly, $\rho^{(2)}_{cv\blk}=0$) we obtain
\begin{eqnarray}
F^{(1p-NI)}_{cc'\blk} &=&
\ble(t)\cdot\Bigg[\sum_{v}\left(\mathbf{d}_{cv\blk}\rho_{vc'\blk}^{(1)}-\rho_{cv\blk}^{(1)}\mathbf{d}_{vc'\blk}\right)\Bigg] \, .  \label{eq:dm_1st_order_probe_cc}   \\
F^{(1p-NI)}_{vv'\blk} &=&
\ble(t)\cdot\Bigg[\sum_{c}\left(\mathbf{d}_{vc\blk}\rho_{cv'\blk}^{(1)}-\rho_{vc\blk}^{(1)}\mathbf{d}_{cv'\blk}\right)\Bigg] \, .  \label{eq:dm_1st_order_probe_vv}   \\
F^{(1p-NI)}_{cv\blk} &=&
\ble(t)\cdot\Bigg[
\sum_{c'}\left(-\rho_{cc'\blk}^{(2)}\mathbf{d}_{c'v\blk}\right)+
\sum_{v'}\left(\mathbf{d}_{cv'\blk}\rho_{v'v\blk}^{(0+2)}\right)\Bigg] \, ,  \label{eq:dm_1st_order_probe_cv} 
\end{eqnarray}
All leading order terms are included, thus we expect the intra-band dipoles not to have an important role in the coupling with the external field. The resulting source term have contributions from first order terms in the pump pulse, at variance with the dephased case, where only second order terms are present. The first order terms affects only the $cc'$ and $cv'$ channels. However, this may lead to oscillations in the $cv$ energy range, since $\rho_{cv\blk}$ enters in such source terms. Thus intra-band dipoles may have an important role in the definition of the polarization. This is why a complete discussion on TrAbs signal may need to account for intra-band dipoles in the excitonic case.

\subsubsection{HSEX approximation}

We are interested to the HSEX self-energy to linear order in the probe pulse,
with a first term coming from $[\Delta\Sigma^{(1+2)},\rho^{(1p)}]_{nm\blk}$ 
and the second from $[\Delta\Sigma^{(1p)},\rho^{(0+1+2)}]_{nm\blk}$
\begin{multline}
G^{(1p)}_{nm\blk}= 
 K^{HSEX}_{nl'\blk,n'm'\blk'}\rho^{(1+2)}_{n'm'\blk'}\rho^{(1p)}_{l'm\blk}
-\rho^{(1p)}_{nl'\blk}K^{HSEX}_{l'm\blk,n'm'\blk'}\rho^{(1+2)}_{n'm'\blk'}+ \\
+K^{HSEX}_{nl'\blk,n'm'\blk'}\rho^{(1p)}_{n'm'\blk'}\rho^{(0+1+2)}_{l'm\blk}
-\rho^{(0+1+2)}_{nl'\blk}K^{HSEX}_{l'm\blk,n'm'\blk'}\rho^{(1p)}_{n'm'\blk'}
\end{multline}
where here $\rho^{(0+1+2)}$ is the density matrix up to second order in the pump, while $\rho^{(1p)}$ is the density matrix also including the first order correction due to the probe.
Taking the $cv$ channel,  the second term can be re-written as $\tilde{K}^{P}_{nm\blk,n'm'\blk'}\rho^{(1p)}_{n'm'\blk'}$.
The first term instead contains $(\Delta\Sigma^{(1+2)}_{cc\blk}-\Delta\Sigma^{(1+2)}_{vv\blk})\rho^{(1p)}_{cv\blk}
\equiv\Delta\tilde{\Sigma}^P_{cv\blk}\rho^{(1p)}_{cv\blk}$ which can be interpreted as a renormalization of the QP band structure,
plus $\Delta\Sigma^{(1+2)}_{cv\blk}(\rho^{(1p)}_{vv\blk}-\rho^{(1p)}_{cc\blk})$
which points to the correlation between diagonal and off diagonal terms and the role of intraband dipoles. Neglecting this last term we have, w.r.t. the equilibrium case the change induced by the pump
\begin{equation}
\left(\Delta\epsilon_{cv\blk}\delta_{v,\vt}\delta_{c,\ct}\delta(\blk-\blkt)+(f^{eq}_{v\blk}-f^{eq}_{c\blk})K^{HSEX}_{cv\blk,\ct\vt\blkt}\right)\rho^{(1p)}_{\ct\vt\blkt}
\rightarrow
\left((\Delta\epsilon_{cv\blk}+\Delta\tilde{\Sigma}^P_{cv\blk})\delta_{v,\vt}\delta_{c,\ct}\delta(\blk-\blkt)+\tilde{K}^{P}_{cv\blk,\ct\vt\blkt}\right)\rho^{(1p)}_{\ct\vt\blkt}
\end{equation}

\subsection{On the role of intraband dipoles}

\begin{figure}[h]
	\begin{center}
		\includegraphics[width=0.325\textwidth]{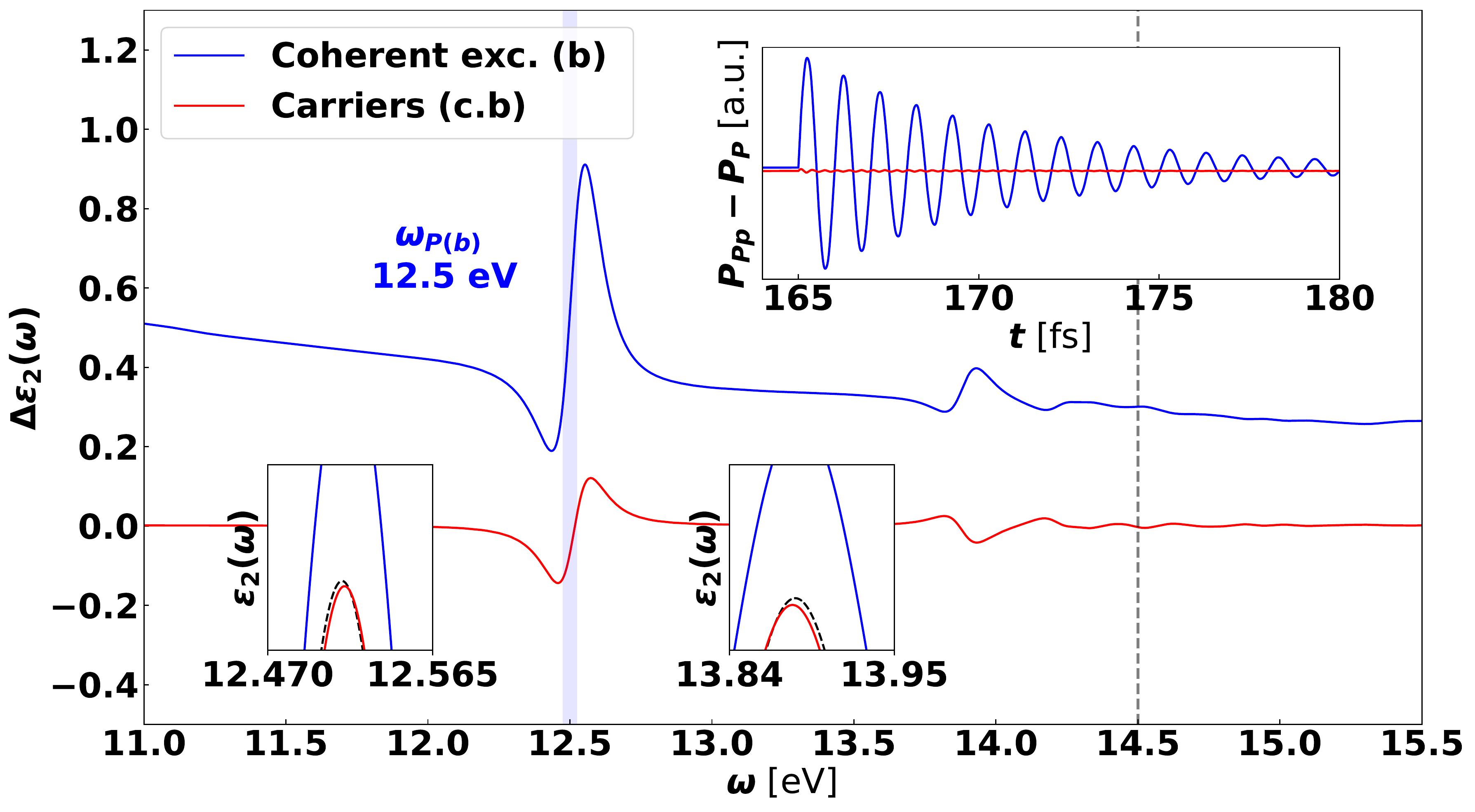}
		\includegraphics[width=0.325\textwidth]{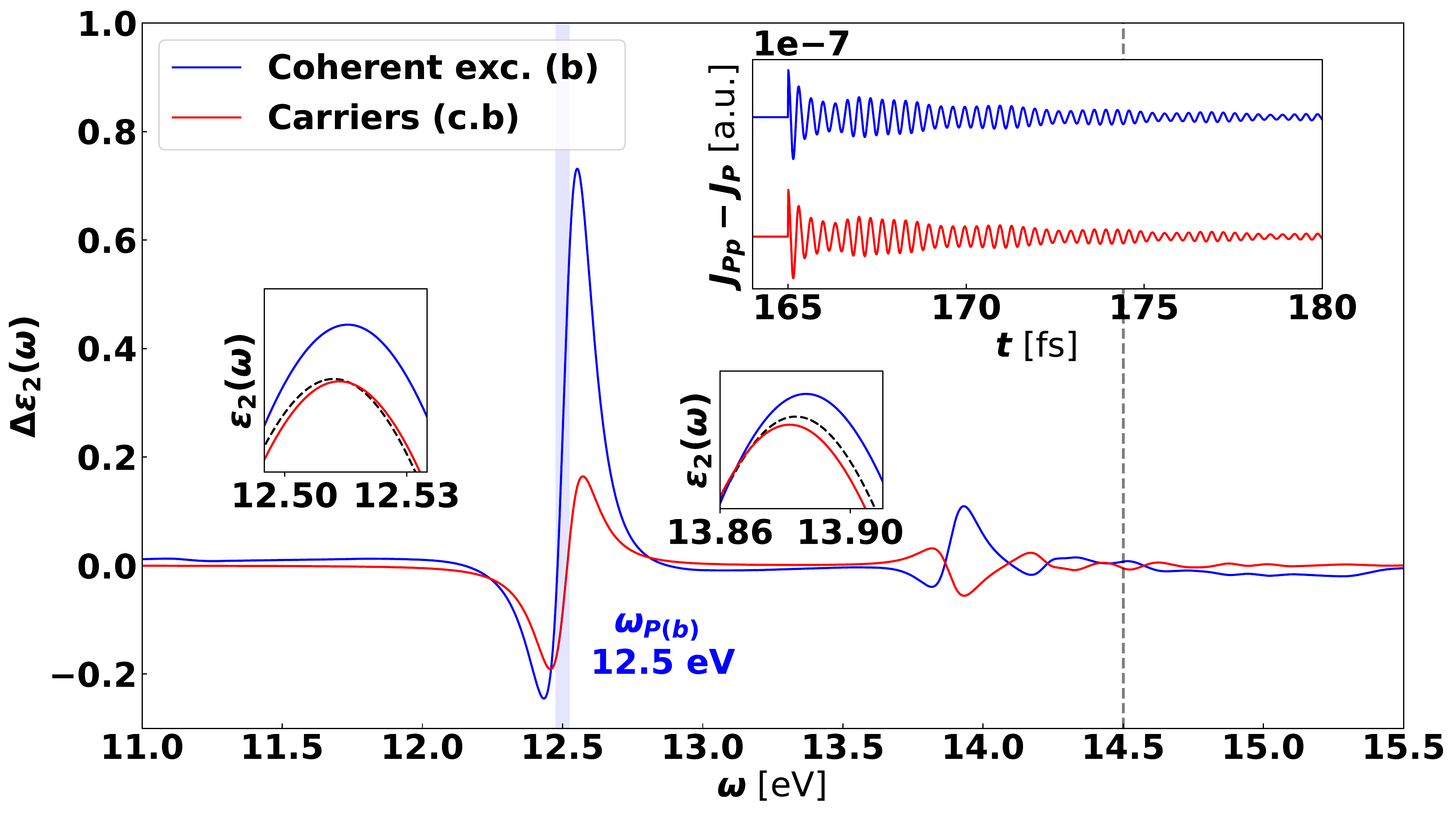}
		\includegraphics[width=0.325\textwidth]{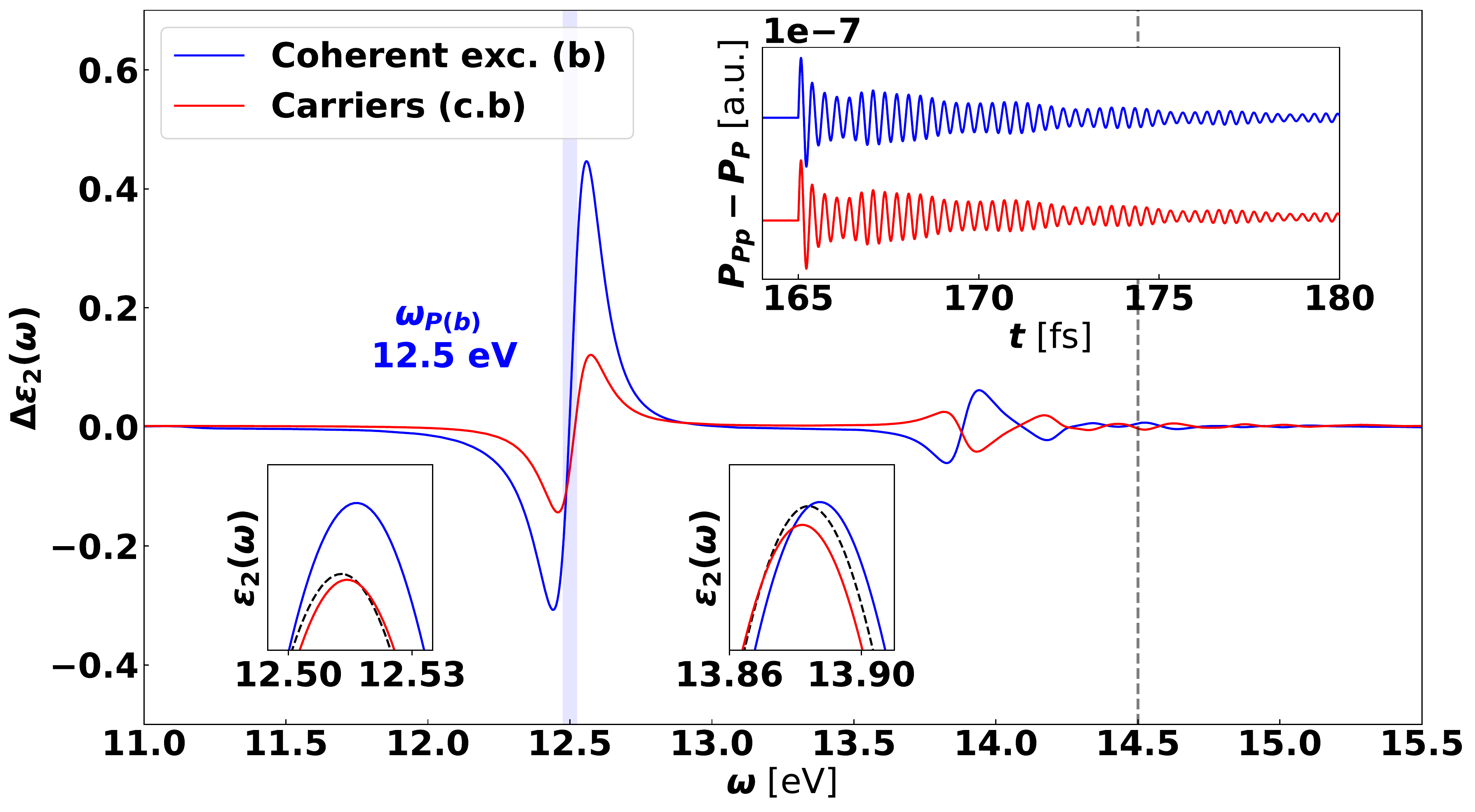}
	\end{center}
	\caption{Transient Absorption for LiF shown for the system driven with a pulse resonant to the excitonic peak. In all cases the states are generate with a TD-HSEX propagation up to 160 fs. Starting from the generated density the system is then probed with a delta like pulse. In the time range 160-260 fs, the propagation in the first and second panel from the left is done including intra-band dipoles. In the first panel the spectrum is reconstructed starting from the polarization, in the second starting from the current. Finally in the third panel the propagation is done setting to zero the intra-band dipoles, and then computing the spectrum from the polarization. The probe induced polarization $\blP^{(i)}(t)=\blP_{Pp}(t)-\blP_{P}(t)$ is shown in the inset.
	The blue shaded line marks the laser pump frequency, the black dashed line the position of the quasi-particle gap. The extra insets show a zoom of $\varepsilon^{(eq)}_2(\omega)$ (black line) and $\varepsilon^{(i)}_2(\omega)$ around the two main excitonic peaks.}
	\label{fig:TR-ABS_intraband}
\end{figure}

To verify the role of intraband dipoles, we compare in Fig.~\ref{fig:TR-ABS_intraband} the transient absorption signal using three different schemes which deal in different ways with the intra-band dipoles. The first panel shows the transient absorption generated from the Fourier transform of ${\blP_{Pp}-\blP_P}$ including intra-band diploles. Since such dipoles are ill defined for small frequencies the non-equilibrium signal in presence of the coherent excitonic state has a non physical behavior. This appears as a $1/\omega$ decay in the detected energy range of the plot, which is due to a non-physical pole appearing at $\omega_X\approx 4.11$~eV (not shown)  with huge intensity ($\epsilon(\omega_X)\approx 200$). The existence of such pole is responsible for the behavior of the polarization, which is dominated by a characteristic frequency in the first panel.  If instead the spectrum is reconstructed starting from the current ${\blJ_{Pp}-\blJ_P}$ (central panel), such pole disappears. The current based approach is defined in terms of the expectation value of the velocity operator and the intra--band dipoles are well defined in such case. However the velocity operator should in principle contain also the commutator with the HSEX self-energy which is neglected here. As a last step we compute again the transient absorption starting from the ${\blP_{Pp}-\blP_P}$, but setting to zero all intra-band dipoles. The signal is quite similar to the one obtained in the current based approach. 
In particular in both cases the differences between the coherent excitonic case and the non coherent populations discussed in the main text remain the same.
Further checks on the role of the intra--band dipoles could be achieved via the NEQ Berry Phase formulation.

\subsection{Tr-Abs signal in the continuum}  \label{ssec:Abs_continuum}

\begin{figure}[h]
	\begin{center}
		\includegraphics[width=0.325\textwidth]{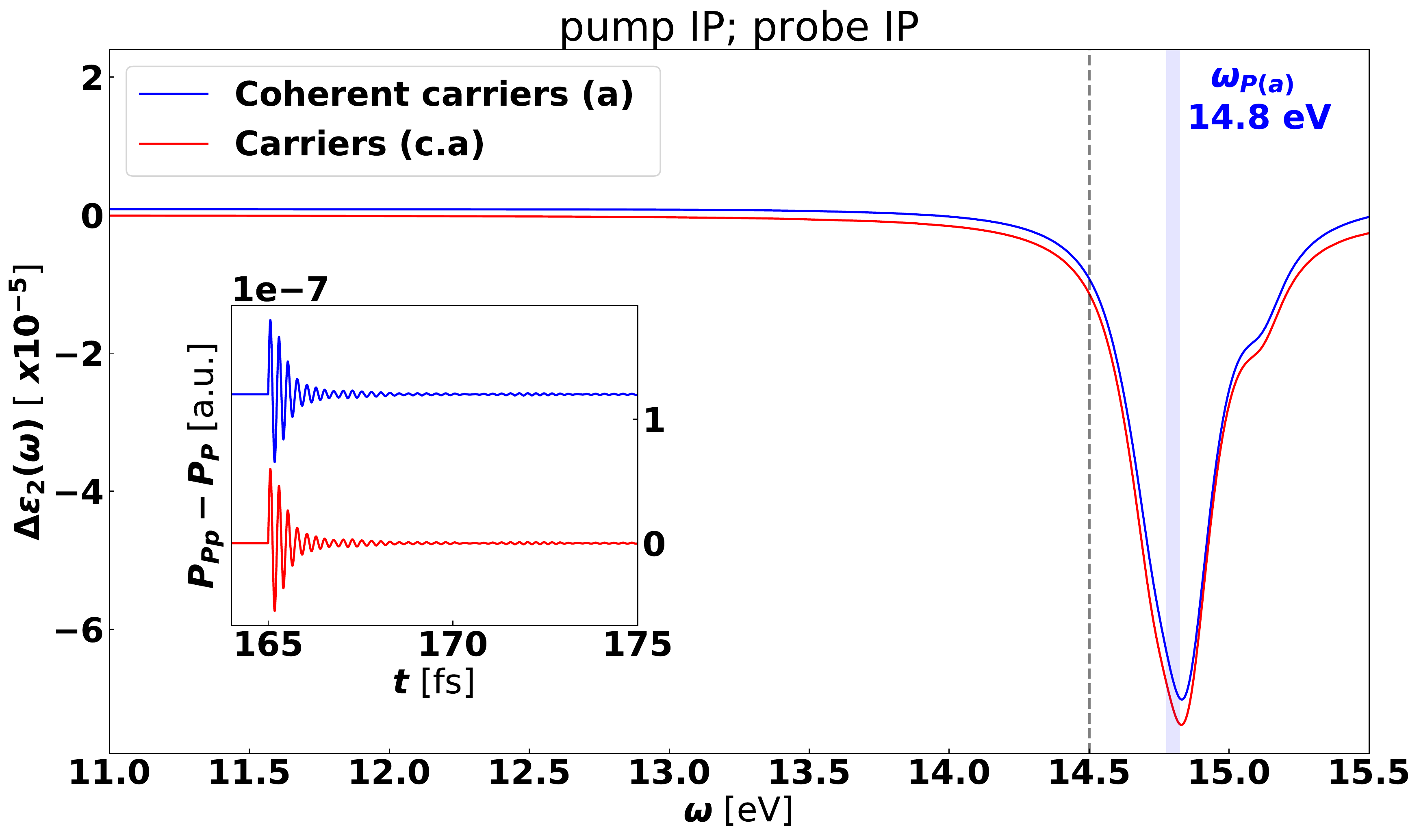}
		\includegraphics[width=0.325\textwidth]{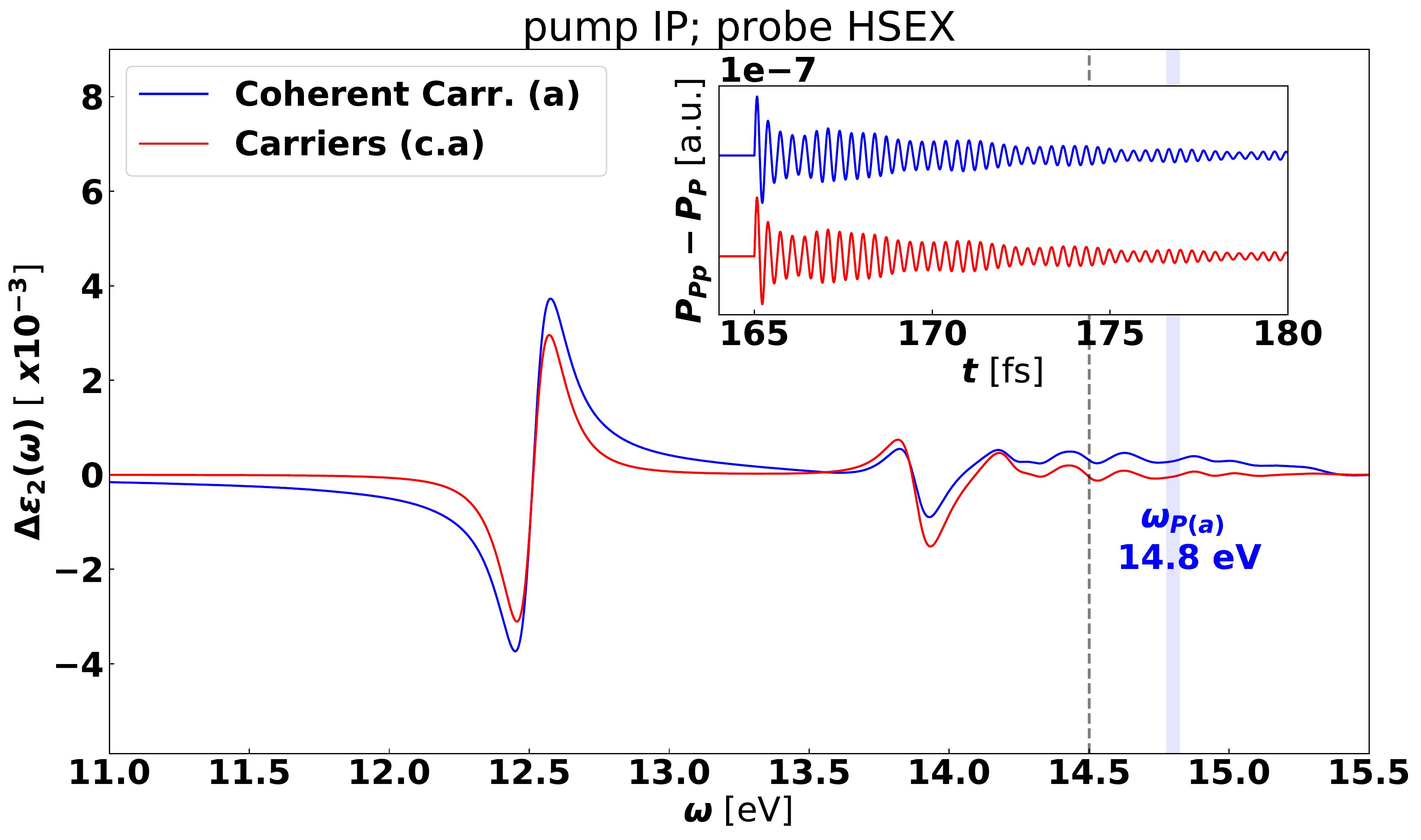}
		\includegraphics[width=0.325\textwidth]{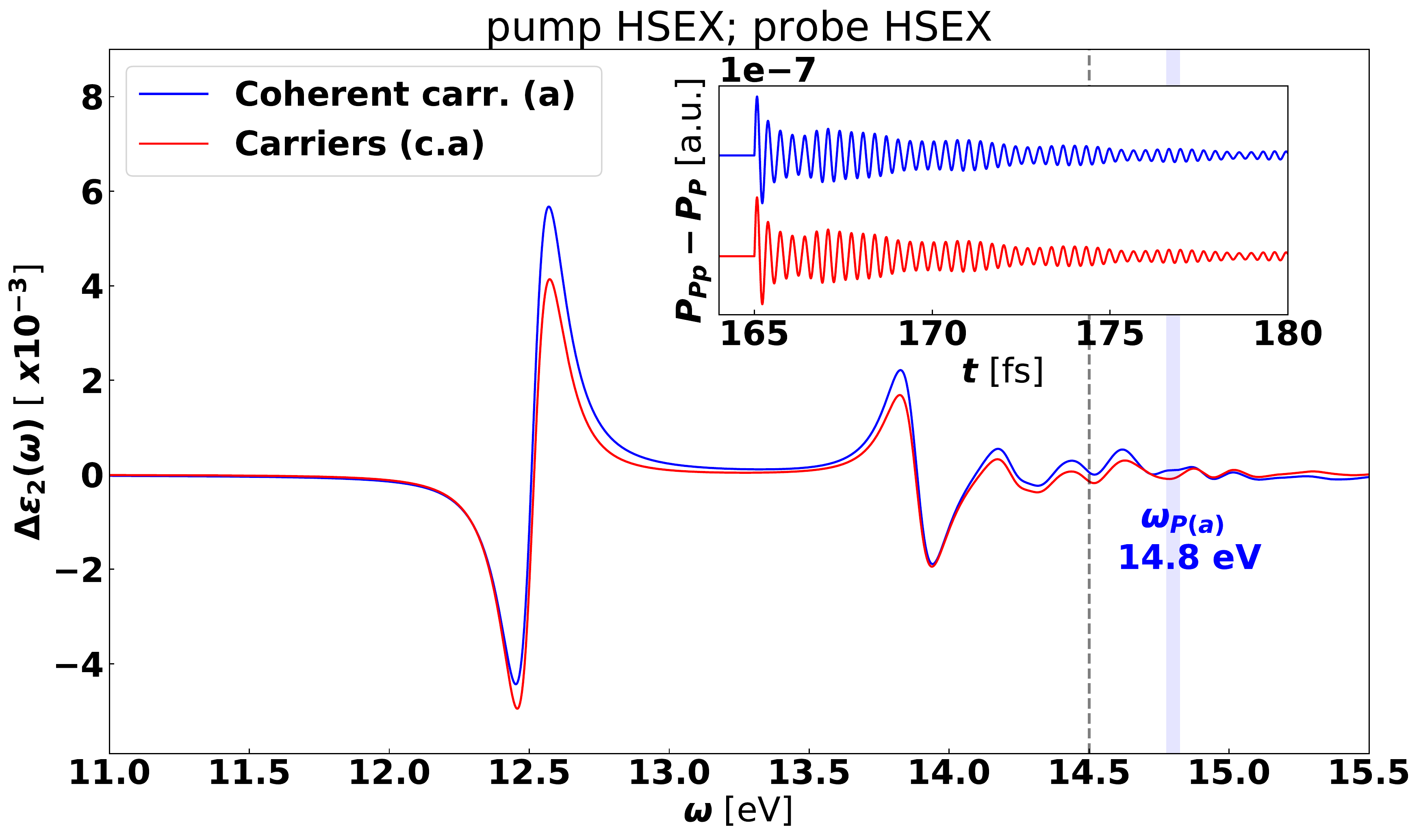}
	\end{center}
	\caption{Transient Absorption for LiF shown for the system driven in the continuum. In both cases the states are generate with a TD-IP propagation up to 160 fs. Starting from the generated density the system is then probed with a delta like pulse propagating either the TD-IP EOM (left panel) and the TD-HSEX EOM (right panel). The response is completly different, showing that excitonic effects cannot be neglected in the probing step. In both cases the role of coherences is negligible.}
	\label{fig:TR-ABS}
\end{figure}

In this section we discuss in detail the idea of using an IP propagation for the generation of the coherent carriers state.
To this end we compare the TrAbs obtained with IP pump - IP probe, IP pump - HSEX probe, and HSEX pump - HSEX probe.
In the IP - IP case the signal is due to Pauli blocking or bleaching of the transitions, mostly the one at 14.8~eV. Other transitions which share either the conduction band or the valence band with the transition at 14.8~eV are also slightly bleached. All this can be easely interpreted in terms of $\chi^{IP}[f_{n\blk}(\tau)](\omega)$ with $f_{n\blk}=\rho_{nn\blk}$.
The signal completly neglect excitonic effects in the pump process. This is not a good approximation, as discussed in the main text. Instead, the IP - HSEX signal and the HSEX - HSEX signal are quite similar (the main difference is the overall magnitude), confirming the the IP approximation for the generation of the state is a reasonable approximation.

\subsection{On the structure of the response function}

I discuss here the resonant and non resonat term of the response function and the relation between these two terms and the absorption and emission channel. I first start writing the IP response function for a cold semiconductor, i.e. with $f_{v\blk}=1$ and $f_{c\blk}=0$ for valence and conduction respectively.
The expression of a general response function contains only $v\rightarrow c$ transitions:
\begin{equation}
\chi^{AB,eq}(\w)=\sum_{cv\blk}\frac{(\langle v\blk|\hat{A}|c\blk\rangle\langle c\blk|\hat{B}|v\blk\rangle)f_{v\blk}(1-f_{c\blk})}{\w-(\e_{c\blk}-\e_{v\blk})+i\eta}
-\frac{(\langle v\blk|\hat{B}|c\blk\rangle\langle c\blk|\hat{A}|v\blk\rangle)f_{v\blk}(1-f_{c\blk})}{\w+(\e_{c\blk}-\e_{v\blk})+i\eta}.
\label{eq:res_ares_IP}
\end{equation}
The first term, or resonant term, describes absorption and has poles for $\omega>0$. The second term, or anti-resonant term, would describe emission, but at equilibrium has poles only for $\omega<0$, i.e. no emission takes place from the ground state.

If we now consider the situation where electrons are injected from valence to conduction, i.e. $f_{v\blk}<1$ and $f_{c\blk}>0$ we add extra terms which contain new $c\rightarrow v$ transitions. These can be constructed by swapping $c\leftrightarrow v$ in the previous expression (I neglect for the present discussion $c\rightarrow c'$ and $v\rightarrow v'$) transitions
\begin{equation}
\chi^{AB,emiss}(\w)=\sum_{cv\blk}\frac{(\langle c\blk|\hat{A}|v\blk\rangle\langle v\blk|\hat{B}|c\blk\rangle)f_{c\blk}(1-f_{v\blk})}{\w-(\e_{v\blk}-\e_{c\blk})+i\eta}
-\frac{(\langle c\blk|\hat{B}|v\blk\rangle\langle v\blk|\hat{A}|c\blk\rangle)f_{c\blk}(1-f_{v\blk})}{\w+(\e_{v\blk}-\e_{c\blk})+i\eta}.
\label{eq:res_ares_IP_neq_corr}
\end{equation}
In this new terms the resonant term has poles for $\omega<0$, i.e. there is no extra contribution to the absorption from these transitions, while the anti-resonant term has now poles for $\omega>0$, and indeed emission can take place from these new term.
Considering only the $\omega>0$ region we thus have, for valence to conduction transitions
\begin{equation}
\chi^{AB,\w>0}(\w)=\sum_{cv\blk}\frac{(\langle v\blk|\hat{A}|c\blk\rangle\langle c\blk|\hat{B}|v\blk\rangle)f_{v\blk}(1-f_{c\blk})}{\w-(\e_{c\blk}-\e_{v\blk})+i\eta}
-\frac{(\langle c\blk|\hat{B}|v\blk\rangle\langle v\blk|\hat{A}|c\blk\rangle)f_{c\blk}(1-f_{v\blk})}{\w-(\e_{c\blk}-\e_{v\blk})+i\eta}.
\label{eq:res_ares_IP_all}
\end{equation}
We see that, compared to equilibrium where $f_{v\blk}(1-f_{c\blk})=1$ we have a reduction of absorption due to Pauli blocking, since now $f_{v\blk}(1-f_{c\blk})<1$ and, moreover a further reduction the peak intensity due to the second term with $f_{c\blk}(1-f_{v\blk})>0$ which is multiplied by $-1$.
The overall weight of the transition if $(f_{v\blk}-f_{c\blk})$, as it should.

The same kind of structure can be constructed from the BSE propagator and leads the correlated version of eq.~\eqref{eq:res_ares_IP_all}, i.e. eq.~(6) in the main text. However in the bosonic case, as discussed, the absorption is not reduced but enhanced and this exactly compensate the emission term. Notice that, similarly to the fact that here $c\rightarrow c'$ and $v\rightarrow v'$ transitions are neglected here, in the main text $\omega_{\lambda\blq}\rightarrow\omega_{\lambda\blq'}$ transitions are neglected.

Let me add few final comments here. The way the bosonic occupations enter eq.~(6) in the main text, with $N_{\lambda\blq}+1$ in one term, and $N_{\lambda\blq}$ in the other, is the ``standard expression'', which, for the electron--magnetic field, is associated to spontaneous emission, stimulated emission and absorption of photons. Here the names are swapped. We call "absorption term" the one which contains the $N_{\lambda\blq}+1$ factor, since we keep referring to the absorption of photons. Indeed $N_{\lambda\blq}+1$ factor describes the creation of excitons, i.e. it could be called exciton emission term, and similarly the second term could be called exciton absorption term, since it describes the destruction of an exciton.
In the electron--magnetic case the +1 term is associated to a ``spontaneous'' process since we start from the electronic system, which interacts with the photon field, in an excited state.
Similarly the emission of excitons is spontaneous (i.e. it does not need the existence of a coherent excitonic source to take place) if we start from the photonic system, which interacts with the ``exciton field'', in an excited state. In this jargon there would be available photons (i.e. the laser pulse) which can ``decay'' by creating excitons to conserve the total energy of the system. There is one case however, where the existence of photon populations is not needed: when $\omega_{\lambda}<0$. This is the case where the EI is spontaneously created. In the present manuscript I'm considering the case with $\omega_{\lambda}>0$ and the NEQ-EI is created via the presence of a laser pulse.

\bibliography{manuscript}